\documentclass[aps,prb,twocolumn,showpacs,preprintnumbers]{revtex4}
\usepackage{amsmath}
\usepackage{graphicx}
\usepackage{soul}
\usepackage{amsfonts}
\usepackage{amssymb}
\usepackage{dcolumn}

\usepackage[T1]{fontenc}
\usepackage[latin9]{inputenc}
\usepackage{booktabs}
\usepackage{amsmath}
\usepackage{esint}
\usepackage{lipsum}

\usepackage{float}

\usepackage[utf8]{luainputenc}
\usepackage{rotating}
\usepackage{stackrel}
\usepackage{esint}

\makeatletter

\begin{document}

\title{An electromagnetic multipole expansion beyond the long-wavelength approximation}

\author{R. Alaee,{$^{*1,2}$} C. Rockstuhl$^{1,3}$ and I. Fernandez-Corbaton$^{3}$ }
\address{$^1$Institute of Theoretical Solid State Physics, Karlsruhe Institute of Technology, 76131 Karlsruhe, Germany}
\address{$^2$Max Planck Institute for the Science of Light, Erlangen 91058, Germany}
\address{$^3$Institute of Nanotechnology, Karlsruhe Institute of Technology, 76021 Karlsruhe, Germany}
\address{$^*$Corresponding author:rasoul.alaee@mpl.mpg.de}

\begin{abstract}
The multipole expansion is a key tool in the study of light-matter interactions. All the information about the radiation of and coupling to electromagnetic fields of a given charge-density distribution is condensed into few numbers: The multipole moments of the source. These numbers are frequently computed with expressions obtained after the long-wavelength approximation. Here, we derive exact expressions for the multipole moments of dynamic sources that resemble in their simplicity their approximate counterparts. We validate our new expressions against analytical results for a spherical source, and then use them to calculate the induced moments for some selected sources with a non-trivial shape. The comparison of the results to those obtained with approximate expressions shows a considerable disagreement even for sources of subwavelength size. Our expressions are relevant for any scientific area dealing with the interaction between the electromagnetic field and material systems.
\end{abstract}
\maketitle

The multipolar decomposition of a given charge-current distribution is taught in every undergraduate course in physics. The resulting set of numbers are called the multipolar moments. They are classified according to their order, i.e. dipoles, quadrupoles etc... For each order, there are electric and magnetic multipolar moments. Each multipolar moment is uniquely connected to a corresponding multipolar field. Their importance stems from the fact that the multipolar moments of a charge-current distribution completely characterize both the radiation of electromagnetic fields by the source, and the coupling of external fields onto it. The multipolar decomposition is important in any scientific area dealing with the interaction between the electromagnetic field and material systems. In particle physics, the multipole moments of the nuclei provide information on the distribution of charges inside the nucleus. In chemistry, the dipole and quadrupolar polarizabilities of a molecule determine most of its properties. In electrical engineering, the multipole expansion is used to quantify the radiation from antennas. And the list goes on.

In this Letter, we present new exact expressions for the multipolar decomposition of an electric charge-current distribution. They provide a straightforward path for upgrading analytical and numerical models currently using the long-wavelength approximation. After the upgrade, the models become exact. The expressions that we provide are directly applicable to the many areas where the multipole decomposition of electrical current density distributions is used. For the sake of concreteness, in this article we apply them to a specific field: Nanophotonics.

In nanophotonics, one purpose is to control and manipulate light on the nanoscale. Plasmonic or high-index dielectric nanoparticles are frequently used for this purpose ~\cite{Maier:07,Jahani2016}. The multipole expansion provides insight into several optical phenomena, such as Fano resonances~\cite{Lukyanchuk:10,Miroshnichenko:10}, electromagnetically-induced-transparency~\cite{Chiam:09}, directional light emission~\cite{Hancu:13,Fu:13,Coenen:14,AlaeeKerker:15}, manipulating and controlling spontaneous emission~\cite {Rogobete:07,Zambrana:2015,Doeleman2016}, light perfect absorption~\cite{Landy:08,Alaee:15,Alaee:16}, electromagnetic cloaking~\cite{Alu2008,Alu2009}, and optical (pulling, pushing, and lateral) forces~\cite{Barton1989,Vesperinas:11,chen2011,rodriguez2015lateral,Rahimzadegan2016}. In all these cases, an external field induces displacement or conductive currents into the samples. These induced currents are the source of the scattered field. But: How can we calculate the multipole moments of these induced current distributions?

Exact expressions exists and can be found in standard textbooks, e.g. Eq.~(7.20) in Ref.~\onlinecite{Walecka2004} or Eq.~(9.165) in Ref.~\onlinecite{Jackson:98} (without the magnetization current therein) and a new formulation have been recently derived in Ref.~\onlinecite{grahn2012}. However, up to now they are not frequently used in the literature. One reason for this may be their complexity, i.e. they feature differential operators like the curl and/or vector spherical harmonics. Instead, a long-wavelength approximation that considerably simplifies the expressions is very often used. Their integrands contain algebraic functions of the coordinate and current density vectors. Moreover, the approximate expressions resemble those for the multipole moments derived in the context of electro-statics and magneto-statics. To set a starting point, these expressions are documented in Table~\ref{tabel:ME_approx}. The so-called toroidal moments are also included in these expression as the second term in the electric multipole moments~\cite{FerCor2015b}.

Let us investigate the range of validity of the expressions in Table~\ref{tabel:ME_approx} by comparing them with Mie theory. In Mie theory, the solution for the scattering of a plane wave by a sphere is obtained without \textit{any approximation}, i.e.  it is valid for \textit{any wavelength and size of the sphere}. For example, Mie theory allows to compute the individual contributions of each induced electric and magnetic multipole moment to the total scattering cross-section. We will compare those exact individual contributions to the ones obtained using the formulas in Table~\ref{tabel:ME_approx}.

\begin{table*}
\caption{Multipole moments in \textit{long-wavelength} approximation; electric dipole moment (ED, i.e. $p_\alpha$), magnetic dipole moment (MD, i.e. $m_\alpha$), electric quadrupole moment (EQ, i.e. $Q_{\alpha\beta}^{e}$) and magnetic quadrupole moment (MQ, i.e. $Q_{\alpha\beta}^{m}$) where $\alpha,\beta = x,y,z$.\label{tabel:ME_approx}}

\centering{}%
\begin{tabular}{cll}
 &  & \tabularnewline
\midrule
\midrule
\begin{tabular}{c}
\tabularnewline
$\mathrm{ED:}$\tabularnewline
\tabularnewline
\end{tabular} & %
\begin{tabular}{l}
\tabularnewline
$p_{\alpha}\approx-\frac{1}{i\omega}\left\{ \int d^{3}\mathbf{r}J_{\alpha}^{\omega}+\frac{k^{2}}{10}\int d^{3}\mathbf{r}\left[\left(\mathbf{r}\cdot\mathbf{J}_{\omega}\right)r_{\alpha}-2r^{2}J_{\alpha}^{\omega}\right]\right\} $\tabularnewline
\tabularnewline
\end{tabular} & $\left(\mathrm{T1-1}\right)$\tabularnewline
\midrule
\begin{tabular}{c}
\tabularnewline
$\mathrm{MD:}$\tabularnewline
\tabularnewline
\end{tabular} & %
\begin{tabular}{l}
\tabularnewline
$m_{\alpha}\approx\frac{1}{2}\int d^{3}\mathbf{r}\left(\mathbf{r}\times\mathbf{J_{\omega}}\right)_{\alpha}$\tabularnewline
\tabularnewline
\end{tabular} & $\mathrm{\left(T1-2\right)}$\tabularnewline
\midrule
\begin{tabular}{c}
\tabularnewline
$\mathrm{EQ:}$\tabularnewline
\tabularnewline
\end{tabular} & %
\begin{tabular}{l}
\tabularnewline
$\begin{alignedat}{1}Q_{\alpha\beta}^{\mathrm{e}} & \approx-\frac{1}{i\omega}\left\{ \int d^{3}\mathbf{r}\left[3\left(r_{\beta}J_{\alpha}^{\omega}+r_{\alpha}J_{\beta}^{\omega}\right)-2\left(\mathbf{r}\cdot\mathbf{J}_{\omega}\right)\delta_{\alpha\beta}\right]\right.\\
 & \left.+\frac{k^{2}}{14}\int d^{3}\mathbf{r}\left[4r_{\alpha}r_{\beta}\left(\mathbf{r}\cdot\mathbf{J}_{\omega}\right)-5r^{2}\left(r_{\alpha}J_{\beta}+r_{\beta}J_{\alpha}\right)+2r^{2}\left(\mathbf{r}\cdot\mathbf{J}_{\omega}\right)\delta_{\alpha\beta}\right]\right\}
\end{alignedat}
$\tabularnewline
\tabularnewline
\end{tabular} & $\left(\mathrm{T1}-3\right)$\tabularnewline
\midrule
$\mathrm{MQ:}$ & %
\begin{tabular}{l}
\tabularnewline
$Q{}_{\alpha\beta}^{m}\approx\int d^{3}\mathbf{r}\left\{ r_{\alpha}\left(\mathbf{r}\times\mathbf{J_{\omega}}\right)_{\beta}+r_{\beta}\left(\mathbf{r}\times\mathbf{J_{\omega}}\right)_{\alpha}\right\} $\tabularnewline
\tabularnewline
\end{tabular} & $\left(\mathrm{T1-4}\right)$\tabularnewline
\bottomrule
\end{tabular}
\end{table*}

\begin{figure*}
\centering
\includegraphics[width=0.98\textwidth]{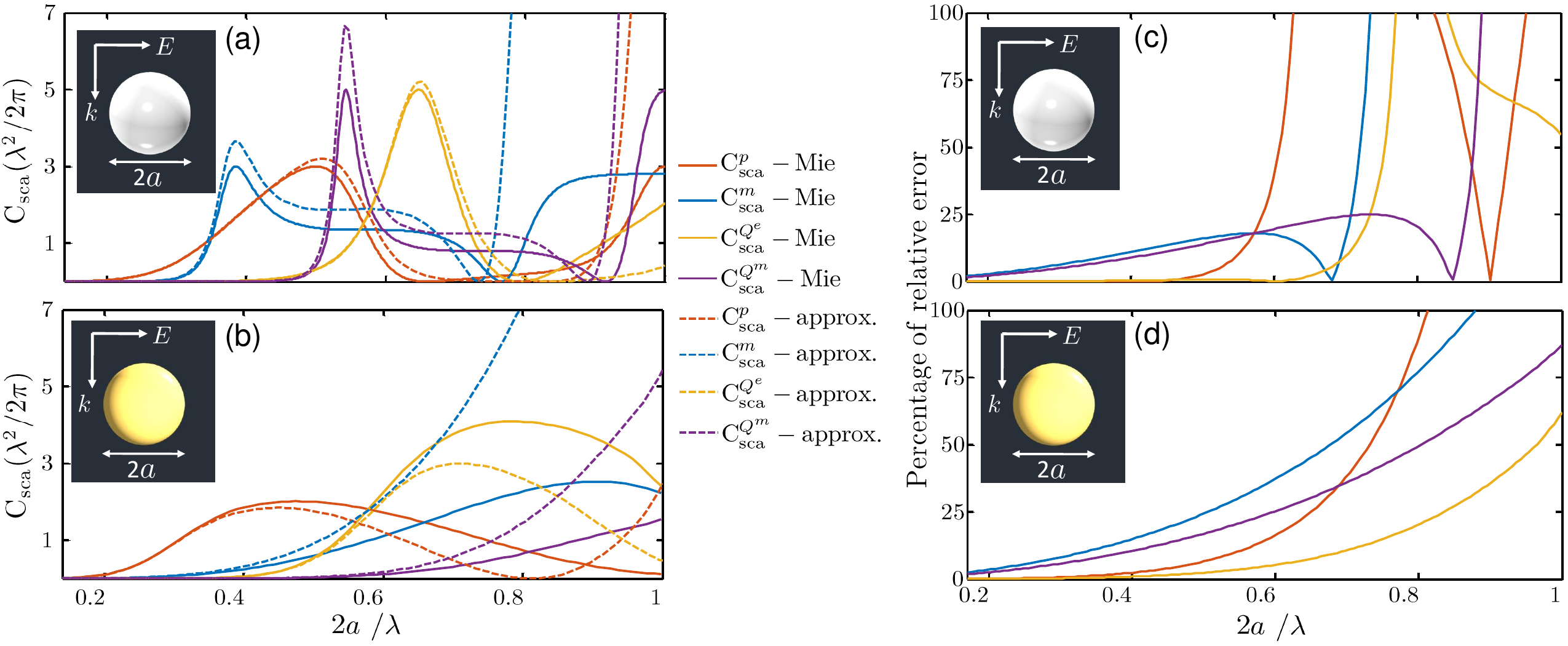}
\caption{Contribution of each multipole moment to the scattering cross section calculated with Mie theory and calculated with the approximate expressions (Table~\ref{tabel:ME_approx}): a) For a dielectric sphere as a function of the particle's size parameter $2a/\lambda$. b) For a gold sphere with a fixed radius of $a$=250~nm. c) and d) Relative error between the multipole moments calculated with the Mie theory and calculated with the approximate expressions. Note that the contribution of each multipole moment to the scattering cross section is normalized to $\lambda^{2}/2\pi$. For spherical particle, there is a universal limit for each multipole, i.e. $\left(2j+1\right)\lambda^{2}/2\pi$. For example, for a dipolar particle (i.e. j=1), the maximum cross section is $3\lambda^{2}/2\pi$~\cite{Ruan:10,Rahimzadegan2016}.}\label{fig:Mie_Approx}
\end{figure*}

We consider a high-index dielectric nanosphere and a gold nanosphere. Both are illuminated with a linearly $x$-polarized plane wave that propagates in the $z$-direction. The induced multipole moments in both cases can be computed using the expressions in Table~\ref{tabel:ME_approx}. The induced electric current density is obtained by using $\mathrm{J}_{\omega}\left(\mathbf{r}\right)=-i\omega\epsilon_{0}\left(\epsilon_{r}-1\right)\mathrm{E_{\omega}}\left(\mathbf{r}\right)$, where $\mathrm{E_{\omega}}\left(\mathbf{r}\right)$ is the electric field distribution, $\epsilon_{0}$ is the permittivity of free space, and $\epsilon_{r}$ is the relative permittivity of the sphere. The permittivity of the dielectric sphere is assumed to be $\epsilon_{r}=2.5^2$. Dispersive material properties as documented in the literature are considered for gold~\cite{Johnson:72}. We assume air as the host medium. We used a numerical finite element solver to obtain the electric field distributions~\cite{multiphysics2012}.

Using the multipole moments, it is easy to obtain the total scattering cross section, i.e. the sum of the contributions from different multipole moments, as~\cite{Jackson:98}:
\begin{eqnarray}
C_{\mathrm{sca}}^{\mathrm{total}} & = & C_{\mathrm{sca}}^{p}+C_{\mathrm{sca}}^{m}+C_{\mathrm{sca}}^{Q^{e}}+C_{\mathrm{sca}}^{Q^{m}}+\cdots\label{eq:C_Sca}\\
 & = & \frac{k^{4}}{6\pi\varepsilon_{0}^{2}\left|\mathbf{E}_{\mathrm{inc}}\right|^{2}}\left[\underset{\alpha}{\sum}\left(\left|p_{\alpha}\right|^{2}+\left|\frac{m_{\alpha}}{c}\right|^{2}\right)+\right.\nonumber \\
 &  & \left.\frac{1}{120}\underset{\alpha\beta}{\sum}\left(\left|kQ_{\alpha\beta}^{e}\right|^{2}+\left|\frac{kQ_{\alpha\beta}^{m}}{c}\right|^{2}\right)+\cdots\right]\nonumber
\end{eqnarray}

where, $p_{\alpha}$, $m_{\alpha}$ are the electric and magnetic
dipole moments, respectively. $Q_{\alpha\beta}^{e}$, $Q_{\alpha\beta}^{m}$
are the electric and magnetic quadrupole moments, respectively. $\ensuremath{|\mathbf{E}_{\mathrm{inc}}|}$ is the amplitude of the incident electric field, $k$ is the wavenumber, and $c$ is the speed of light.

Figure~\ref{fig:Mie_Approx} shows the contribution of each multipole moment to the scattering cross section for a high-index dielectric as well as a gold nanosphere. The results obtained using the approximate expression are compared with those obtained from Mie theory. It can be seen that, upon increasing the $a/\lambda$ ratio, there is a large deviation between the scattering cross section obtained from the expressions in Table~\ref{tabel:ME_approx} and the Mie theory. The relative error between the two approaches is shown in Fig.~\ref{fig:Mie_Approx} (c) and (d). The relative error is more than 100\% for the dielectric sphere at $2a/\lambda\approx0.75$ for both electric and magnetic dipole moments. This large deviation occurs because the expressions in Table~\ref{tabel:ME_approx} are obtained in the long-wavelength approximation~\cite{Jackson:98}, i.e. they are only valid for particles small compared to the wavelength of the incident light (i.e. \textit{$D\ll\lambda$} where $D$ is the biggest dimension of the particle).

Thus, the long-wavelength expressions in Table~\ref{tabel:ME_approx} can \textit{not} be used for large particles (compared to the wavelength). The large deviation observed in Fig.~\ref{fig:Mie_Approx} (c) and (d) for different multipole moments will significantly affect the quantitative prediction of \textit{multipolar interference}, which is the main physical mechanism behind Fano resonances~\cite{Lukyanchuk:10,Miroshnichenko:10}, directional light emission~\cite{Hancu:13,Fu:13,Coenen:14,AlaeeKerker:15}, and light perfect absorption~\cite{Landy:08,Alaee:15}. Moreover, any physical quantity obtained using the multipole moments of Table\ref{tabel:ME_approx}, e.g. absorption/extinction cross section, or optical torque/force, carries a corresponding error. Therefore, the application of the exact expressions for the multipole moments is important since it provides a better understanding of all the highlighted optical phenomena and enables its quantitative prediction.

To improve the situation and indeed to provide error-free expressions, we now derive exact expressions for the induced electric and magnetic multipole moments that are valid for \textit{any wavelength and size} (see Table~\ref{tabel:ME_exact}). They can be used to compute the multipole moments of arbitrarily shaped particles. Our exact expressions for multipole moments are very similar to the well-known expression obtained in long-wavelength approximation(see Table~\ref{tabel:ME_approx}).

Our starting point are the hybrid integrals in Fourier and coordinate space in Eq.~14 of Ref.~\onlinecite{FerCor2015b} (see the {\em supplementary material}). These integrals are exact expressions for all the multipolar moments of a spatially confined electric current density distribution. They are valid for any size of the distribution. Crucially, the Fourier space part of the integrals does not depend on the current density. The results in Tab. \ref{tabel:ME_exact} are obtained after carrying out the Fourier space integrals for the electric and magnetic dipolar and quadrupolar orders (see the {\em supplementary material}). Our results have two main advantages with respect to other exact expressions~\cite{Jackson:98,Walecka2004,grahn2012}. One is that our formulas are simpler: The previously existing expressions contain differential operators and/or vector spherical harmonics inside the integrands, while ours contain algebraic functions of the coordinate and current density vectors, and spherical Bessel functions. The other advantage is that the previous expressions lack the similarity to their long-wavelength approximations that ours have (compare Tabs. \ref{tabel:ME_approx} and \ref{tabel:ME_exact}). Therefor, our expressions allow a straightforward upgrade of analytical and numerical models using the approximated long-wavelength expressions. After the upgrade, the models become exact.

Basically, any code that has been previously implemented to compute the multipole moments with the approximate expression can be made to be accurate with a marginal change.
\begin{table*}
\protect\caption{Exact multipole moments; electric dipole moment (ED, i.e. $p_\alpha$), magnetic dipole moment (MD, i.e. $m_\alpha$), electric quadrupole moment (EQ, i.e. $Q_{\alpha\beta}^{e}$) and magnetic quadrupole moment (MQ, i.e. $Q_{\alpha\beta}^{m}$) where $\alpha,\beta = x,y,z$. The derivation can be found in the {\em supplementary material}.
\label{tabel:ME_exact}}

\centering{}%
\begin{tabular}{cll}
 &  & \tabularnewline
\midrule
\midrule
\begin{tabular}{c}
\tabularnewline
$\mathrm{ED:}$\tabularnewline
\tabularnewline
\end{tabular} & %
\begin{tabular}{l}
\tabularnewline
$p_{\alpha}=-\frac{1}{i\omega}\left\{ \int d^{3}\mathbf{r}J_{\alpha}^{\omega}j_{0}\left(kr\right)+\frac{k^{2}}{2}\int d^{3}\mathbf{r}\left[3\left(\mathbf{r}\cdot\mathbf{J}_{\omega}\right)r_{\alpha}-r^{2}J_{\alpha}^{\omega}\right]\frac{j_{2}\left(kr\right)}{\left(kr\right)^{2}}\right\} $\tabularnewline
\tabularnewline
\end{tabular} & $\left(\mathrm{T2-1}\right)$\tabularnewline
\midrule
\begin{tabular}{c}
\tabularnewline
$\mathrm{MD:}$\tabularnewline
\tabularnewline
\end{tabular} & %
\begin{tabular}{l}
\tabularnewline
$m_{\alpha}=\frac{3}{2}\int d^{3}\mathbf{r}\left(\mathbf{r}\times\mathbf{J_{\omega}}\right)_{\alpha}\frac{j_{1}\left(kr\right)}{kr}$\tabularnewline
\tabularnewline
\end{tabular} & $\mathrm{\left(T2-2\right)}$\tabularnewline
\midrule
\begin{tabular}{c}
\tabularnewline
$\mathrm{EQ:}$\tabularnewline
\tabularnewline
\end{tabular} & %
\begin{tabular}{l}
\tabularnewline
$\begin{alignedat}{1}Q_{\alpha\beta}^{\mathrm{e}} & =-\frac{3}{i\omega}\left\{ \int d^{3}\mathbf{r}\left[3\left(r_{\beta}J_{\alpha}^{\omega}+r_{\alpha}J_{\beta}^{\omega}\right)-2\left(\mathbf{r}\cdot\mathbf{J}_{\omega}\right)\delta_{\alpha\beta}\right]\frac{j_{1}\left(kr\right)}{kr}\right.\\
 & \left.+2k^{2}\int d^{3}\mathbf{r}\left[5r_{\alpha}r_{\beta}\left(\mathbf{r}\cdot\mathbf{J}_{\omega}\right)-\left(r_{\alpha}J_{\beta}+r_{\beta}J_{\alpha}\right)r^{2}-r^{2}\left(\mathbf{r}\cdot\mathbf{J}_{\omega}\right)\delta_{\alpha\beta}\right]\frac{j_{3}\left(kr\right)}{\left(kr\right)^{3}}\right\}
\end{alignedat}
$\tabularnewline
\tabularnewline
\end{tabular} & $\left(\mathrm{T2}-3\right)$\tabularnewline
\midrule
$\mathrm{MQ:}$ & %
\begin{tabular}{l}
\tabularnewline
$Q{}_{\alpha\beta}^{m}=15\int d^{3}\mathbf{r}\left\{ r_{\alpha}\left(\mathbf{r}\times\mathbf{J_{\omega}}\right)_{\beta}+r_{\beta}\left(\mathbf{r}\times\mathbf{J_{\omega}}\right)_{\alpha}\right\} \frac{j_{2}\left(kr\right)}{\left(kr\right)^{2}}$\tabularnewline
\tabularnewline
\end{tabular} & $\left(\mathrm{T2-4}\right)$\tabularnewline
\bottomrule
\end{tabular}
\end{table*}

\begin{figure}
\centering
\includegraphics[width=0.48\textwidth]{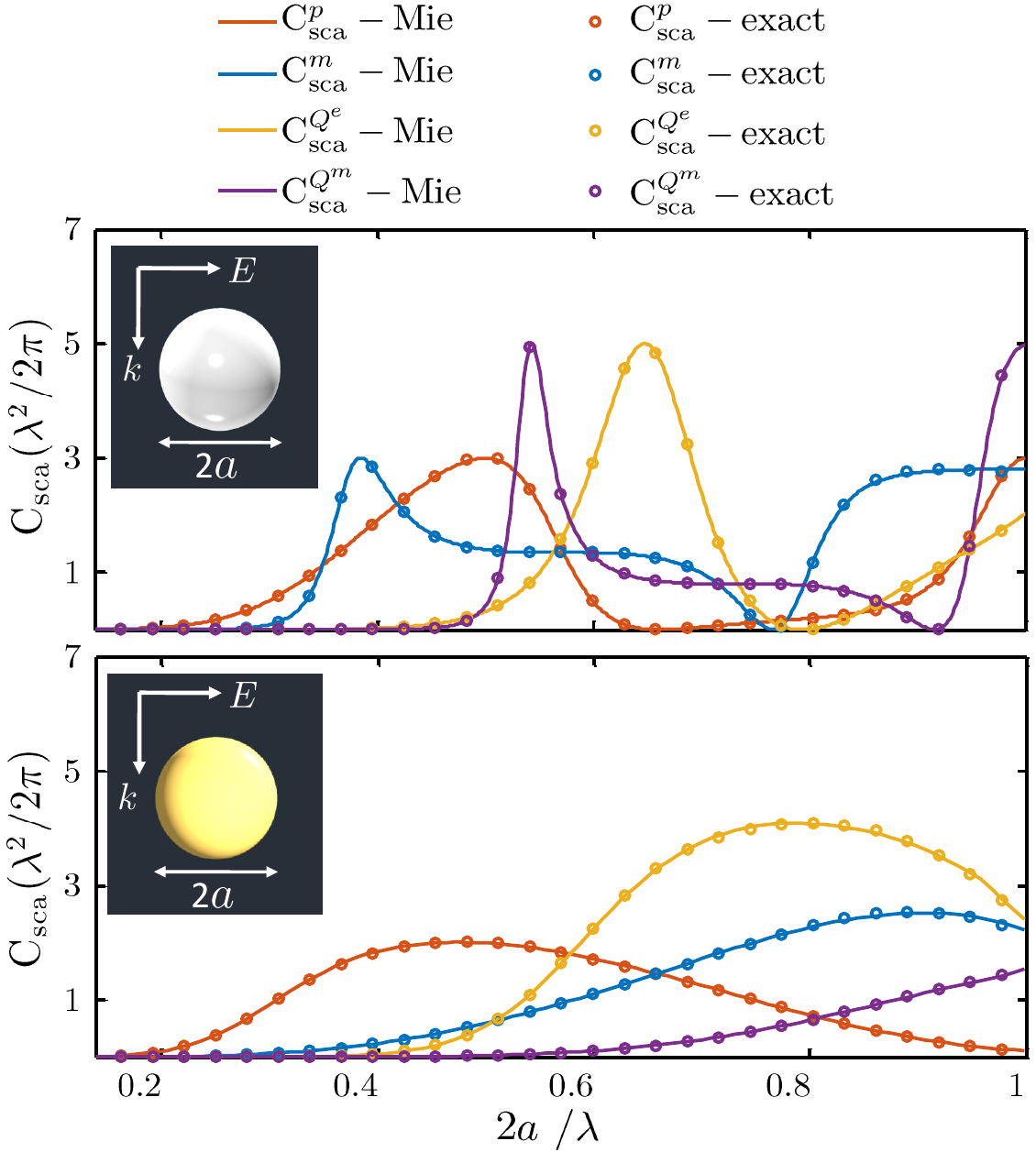}
\caption{Contribution of each multipole moment to the scattering cross section calculated with Mie theory and calculated with the exact expressions (Table~\ref{tabel:ME_exact}). a) For a dielectric sphere with a relative permittivity of $\epsilon_r = 2.5^2$ as a function of the particle's size parameter $2a/\lambda$  b) For a gold sphere with a fixed radius of $a$=250~nm.
}\label{fig:Mie_Exact}
\end{figure}

\begin{figure}
\centering
\includegraphics[width=0.48\textwidth]{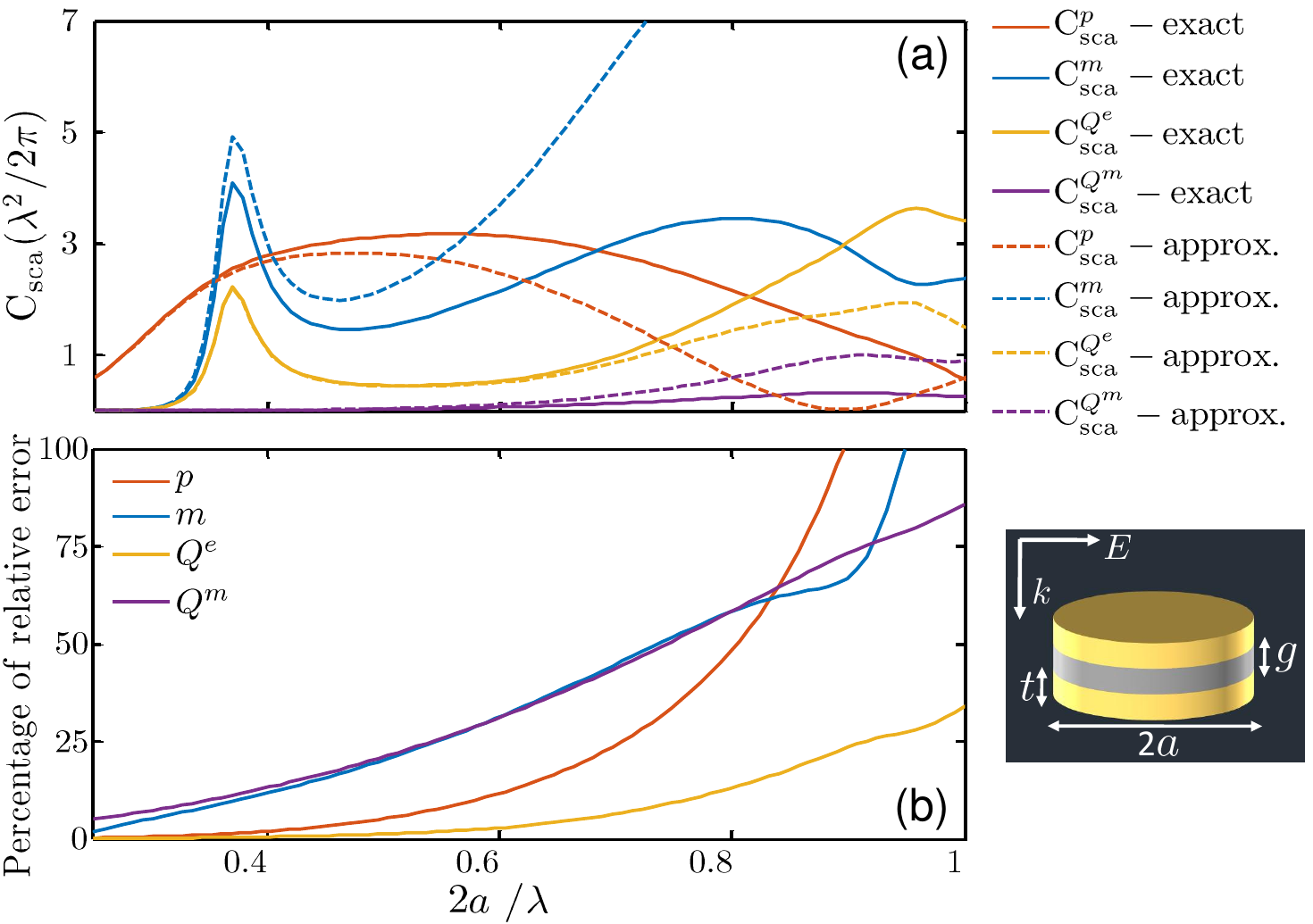}
\caption{a) Contribution of each multipole moment to the scattering cross section calculated with the approximate expressions (Table~\ref{tabel:ME_approx}) and calculated with the exact expressions (Table~\ref{tabel:ME_exact}) for a coupled nanopatch with given geometrical parameters as a function of the wavelength. b) Relative error between the multipole moments calculated with the approximate expression (Table~\ref{tabel:ME_approx}) and calculated with the exact expression (Table~\ref{tabel:ME_exact}).
}\label{fig:CoupledNanopatches}
\end{figure}

In order to show the correctness of the expressions in Table~\ref{tabel:ME_exact}, we compute the contributions of different multipole moments to the scattering cross section and compare them to those obtained with Mie theory. Figure~\ref{fig:Mie_Exact} shows the different contributions as a function of the particle's size parameter $2a/\lambda$ for both the previously considered dielectric and gold spheres. It can be seen that the results from our exact expressions are in excellent agreement with those from Mie theory, irrespective of the particle's size parameter. Indeed, they are indistinguishable up to a numerical noise level.

Up to now, we have considered only spherical particles that could also be studied with Mie theory. We now use the new expressions in Table~\ref{tabel:ME_exact} to calculate the induced moments of a canonical particle made of two coupled nanopatches. Its geometry and the results are shown in Fig.~\ref{fig:CoupledNanopatches}. The coupled nanopatches support a strong electric and magnetic response. The radius and thickness of the coupled disk is assumed to be $a=250$~nm, $t=80$~nm, respectively. The spacer between the two disks is $g=120$~nm. 
It can be seen that there is a significant deviation between the contributions to the scattering cross section from the different multipole moments as predicted by the approximate (Table~\ref{tabel:ME_approx}) and by the exact (Table~\ref{tabel:ME_exact}) expressions. The relative error is shown in Fig.~\ref{fig:CoupledNanopatches} (b). Some of them reach 25\% for a particle size of about half the wavelength.

Finally, there are a few important facts about the expressions shown in Table~\ref{tabel:ME_exact} that are worth highlighting:
\begin{itemize}
\item  The exact multipole moments are valid for any particle's size (i.e. $a/\lambda$) and arbitrarily shaped particles. Note, any physical quantities obtained from the these multipole moments will be exact.
\item There is \textit{no} need to introduce a third family of multipole (i.e. toroidal multipole moments). Our new expressions reveal that toroidal multipole moments are only the higher order terms in the expansion of the electric multipole moments~\cite{FerCor2015c}.
\item The well known approximate multipole moments in Table~\ref{tabel:ME_approx} can be obtained from the expressions in Table~\ref{tabel:ME_exact} by using a long-wavelength approximation. This means that the approximate expression in Table~\ref{tabel:ME_approx} can be easily recovered by making a small argument approximation to the spherical Bessel functions (see the {\em supplementary material}):
\end{itemize}
\begin{eqnarray*}
j_{0}\left(kr\right) & \approx & 1-\left(kr\right)^{2}/6,\\
j_{1}\left(kr\right) & \approx & kr/3,\\
j_{2}\left(kr\right) & \approx & \left(kr\right)^{2}/15.
\end{eqnarray*}

In summary, we have introduced new expressions for multipole moments [Table~\ref{tabel:ME_exact}] which are valid for arbitrarily sized particles of any shape. The well-known long-wavelength expression (Table I) are recovered as the lowest order terms of our new exact expressions (Table II). We have shown the correctness of our expressions by comparing their results with those of Mie theory and obtaining a complete agreement. We are confident that our new exact expressions in Table~\ref{tabel:ME_exact} have the potential to be used in every electrodynamics textbook and actually should be taught in undergraduate courses in physics.

Beyond the particular case of multipolar moments induced by an incident field in a structure, our expressions can be directly applied in the many areas where the multipole decomposition of electrical current density distributions is used.

\section*{Acknowledgements}
The authors warmly thank Dr. Zeinab Mokhtari, Renwen Yu and Burak G\"urlek for their constructive comments and suggestions. We acknowledge the German Science Foundation for support within the project RO 3640/7-1. R.A. would like to acknowledge financial support from the Max Planck Society.

\appendix
\section{Exact expressions for the multipole moments}
Let us start with the hybrid integrals in Fourier and coordinate space
(see our previous work~\cite{FerCor2015b}):
\begin{widetext}
\begin{eqnarray}
\frac{\sqrt{\left(2\pi\right)^{3}}}{4\pi}a_{jm}^{\omega} & = & \stackrel[\bar{l}\bar{m}]{}{\sum}\left(-i\right)^{\bar{l}}\int d\widehat{\mathbf{p}}\mathbf{Z}_{jm}^{\dagger}\left(\widehat{\mathbf{p}}\right)Y_{\bar{l}\overline{m}}\left(\widehat{\mathbf{p}}\right)\int d^{3}\mathbf{r}\mathbf{J_{\omega}\left(r\right)}Y_{\bar{l}\overline{m}}^{*}\left(\widehat{\mathbf{r}}\right)j_{\bar{l}}\left(kr\right),\label{eq:EM}\\
\frac{\sqrt{\left(2\pi\right)^{3}}}{4\pi}b_{jm}^{\omega} & = & \stackrel[\bar{l}\bar{m}]{}{\sum}\left(-i\right)^{\bar{l}}\int d\widehat{\mathbf{p}}\mathbf{X}_{jm}^{\dagger}\left(\widehat{\mathbf{p}}\right)Y_{\bar{l}\overline{m}}\left(\widehat{\mathbf{p}}\right)\int d^{3}\mathbf{r}\mathbf{J_{\omega}\left(r\right)}Y_{\bar{l}\overline{m}}^{*}\left(\widehat{\mathbf{r}}\right)j_{\bar{l}}\left(kr\right),\label{eq:MM}
\end{eqnarray}

where $a_{jm}^{\omega}$ and $b_{jm}^{\omega}$ are exact expressions
for the multipole moments (electric and magnetic, respectively) of
a spatially localized electric current density distribution $\mathbf{J_{\omega}\left(r\right)}$.
These expressions are valid for any size of the current distribution.
As shown in Ref.~\cite{FerCor2015b}, only terms with $\bar{l}=j$
contribute to the $b_{jm}^{\omega}$, whereas $a_{jm}^{\omega}$ has
contributions from both $\bar{l}=j-1$ and $\bar{l}=j+1$. $\mathbf{X}_{jm}\left(\widehat{\mathbf{p}}\right)$
and $\mathbf{Z}_{jm}\left(\widehat{\mathbf{p}}\right)$ are the multipolar
functions in momentum space and defined as

\begin{eqnarray}
\mathbf{X}_{jm}\left(\widehat{\mathbf{p}}\right) & = & \frac{1}{\sqrt{j\left(j+1\right)}}\mathbf{L}Y_{jm}\left(\widehat{\mathbf{p}}\right),\\
\mathbf{Z}_{jm}\left(\widehat{\mathbf{p}}\right) & = & i\widehat{\mathbf{p}}\times\mathbf{X}_{jm}\left(\widehat{\mathbf{p}}\right),
\end{eqnarray}

where $Y_{jm}\left(\widehat{\mathbf{p}}\right)$ is the spherical
harmonics and three components of the vector $\mathbf{L}$ are the
angular momentum operators for scalar function. $\widehat{\mathbf{p}}$
is the angular part of the momentum vector $\mathbf{p}$ ($\left|\mathbf{p}\right|=\frac{\omega}{c}$).

In the following sections, we use Eq.~\ref{eq:EM} and Eq.~\ref{eq:MM}
to derive a simplified exact expressions for the electric multipole
moments in both spherical and Cartesian coordinates. We start by documenting a few auxiliary expressions that will be frequently used at a later stage and consecutively work out afterwards the details for a specific multipolar order of either the electric and magnetic multipole moments. Following the tradition of Jackson, we treat expressions up to the quadrupolar order. If higher orders would be needed, the discussion would be analogous.

\section{Useful expressions in spherical and Cartesian coordinates}

A vector $\mathbf{a}$ in spherical basis defined as

\begin{eqnarray}
\mathbf{a} & = & a_{1}\hat{\mathbf{e}}_{1}+a_{0}\hat{\mathbf{e}}_{0}+a_{-1}\hat{\mathbf{e}}_{-1},\label{eq:A_vec_def}
\end{eqnarray}

with

\begin{eqnarray}
\hat{\mathbf{e}}_{1} & = & -\frac{\hat{\mathbf{x}}+i\mathbf{\hat{y}}}{\sqrt{2}},\nonumber \\
\hat{\mathbf{e}}_{0} & = & \hat{\mathbf{z}},\\
\hat{\mathbf{e}}_{-1} & = & \frac{\hat{\mathbf{x}}-i\mathbf{\hat{y}}}{\sqrt{2}}.\nonumber
\end{eqnarray}

Therefore, the relation between the Cartesian and spherical coordinates
of vector $\mathbf{a}$ reads as:

\begin{eqnarray}
\left[\begin{array}{c}
a_{1}\\
a_{0}\\
a_{-1}
\end{array}\right] & = & \left[\begin{array}{ccc}
-\frac{1}{\sqrt{2}} & \frac{i}{\sqrt{2}} & 0\\
0 & 0 & 1\\
\frac{1}{\sqrt{2}} & \frac{i}{\sqrt{2}} & 0
\end{array}\right]\left[\begin{array}{c}
a_{x}\\
a_{y}\\
a_{z}
\end{array}\right],\label{eq:A_T_CarSph}
\end{eqnarray}

and the cross product of two vector in spherical basis can be defined
as:

\begin{eqnarray}
\mathbf{a}\times\mathbf{b} & = & i\left[\begin{array}{c}
a_{1}b_{0}-a_{0}b_{1}\\
a_{1}b_{-1}-a_{-1}b_{1}\\
a_{0}b_{-1}-a_{-1}b_{0}
\end{array}\right].\label{eq:CrossP_ab}
\end{eqnarray}

Let us now introduce a few useful relations between spherical and
Cartesian coordinates which will be used in the following sections.

The vector $\mathbf{r}$ and $\mathbf{J_{\omega}}$ in spherical basis
can be written as:

\begin{eqnarray}
\mathbf{J_{\omega}} & = & \left[\begin{array}{ccc}
J_{1}^{\omega} & J_{0}^{\omega} & J_{-1}^{\omega}\end{array}\right]^{T},\,\,\,\,\,\,\,\,\,\,\mathbf{r}=\left[\begin{array}{ccc}
r_{1} & r_{0} & r_{-1}\end{array}\right]^{T},\label{eq:J_and_r_def}
\end{eqnarray}

according to Eq.~\ref{eq:A_T_CarSph} for vectors $\mathbf{r}$,
$\mathbf{J_{\omega}}$, $\mathbf{r}\times\mathbf{J_{\omega}}$ the
relation between the spherical and Cartesian coordinates read as:

\begin{equation}
\left[\begin{array}{c}
x\\
y\\
z
\end{array}\right]=\left[\begin{array}{c}
\frac{1}{\sqrt{2}}\left(r_{-1}-r_{1}\right)\\
\frac{1}{\sqrt{2}i}\left(r_{-1}+r_{1}\right)\\
r_{0}
\end{array}\right],\label{eq:r_C2S}
\end{equation}

\begin{equation}
\left[\begin{array}{c}
J_{x}^{\omega}\\
J_{y}^{\omega}\\
J_{z}^{\omega}
\end{array}\right]=\left[\begin{array}{c}
\frac{1}{\sqrt{2}}\left(J_{-1}^{\omega}-J_{1}^{\omega}\right)\\
\frac{1}{\sqrt{2}i}\left(J_{-1}^{\omega}+J_{1}^{\omega}\right)\\
J_{0}^{\omega}
\end{array}\right],\label{eq:J_C2S}
\end{equation}

\begin{eqnarray}
\left[\begin{array}{c}
\left(\mathbf{r}\times\mathbf{J_{\omega}}\right)_{x}\\
\left(\mathbf{r}\times\mathbf{J_{\omega}}\right)_{y}\\
\left(\mathbf{r}\times\mathbf{J_{\omega}}\right)_{z}
\end{array}\right] & = & \left[\begin{array}{c}
\frac{\left(\mathbf{r}\times\mathbf{J_{\omega}}\right)_{-1}-\left(\mathbf{r}\times\mathbf{J_{\omega}}\right)_{1}}{\sqrt{2}}\\
\frac{\left(\mathbf{r}\times\mathbf{J_{\omega}}\right)_{-1}+\left(\mathbf{r}\times\mathbf{J_{\omega}}\right)_{1}}{\sqrt{2}i}\\
\left(\mathbf{r}\times\mathbf{J_{\omega}}\right)_{0}
\end{array}\right].\label{eq:eq:r_cross_J_C2S}
\end{eqnarray}

Using Eq.~\ref{eq:CrossP_ab}, the cross product of vector $\mathbf{r}\times\mathbf{J_{\omega}}$
reads as:

\begin{eqnarray}
\mathbf{r}\times\mathbf{J_{\omega}} & =\left[\begin{array}{c}
\left(\mathbf{r}\times\mathbf{J_{\omega}}\right)_{1}\\
\left(\mathbf{\mathbf{r}}\times\mathbf{J_{\omega}}\right)_{0}\\
\left(\mathbf{r}\times\mathbf{J_{\omega}}\right)_{-1}
\end{array}\right]= & i\left[\begin{array}{c}
J_{0}^{\omega}r_{1}-J_{1}^{\omega}r_{0}\\
J_{-1}^{\omega}r_{1}-J_{1}^{\omega}r_{-1}\\
J_{-1}^{\omega}r_{0}-J_{0}^{\omega}r_{-1}
\end{array}\right],\label{eq:CP_S}
\end{eqnarray}

and the scalar product is

\begin{eqnarray}
\mathbf{r}\cdot\mathbf{J}_{\omega} & = & xJ_{x}^{\omega}+yJ_{y}^{\omega}+zJ_{z}^{\omega}.\label{eq:DP_C}
\end{eqnarray}

Finally, the spherical harmonics has following relations with vector
$\mathbf{r}$ ($\hat{\mathbf{r}}=\frac{\mathbf{r}}{r}$)~\cite{SF_88}:

\begin{eqnarray}
\left[\begin{array}{c}
Y_{11}\\
Y_{10}\\
Y_{1-1}
\end{array}\right] & = & \frac{1}{2}\sqrt{\frac{3}{\pi}}\left[\begin{array}{c}
-\hat{r}_{-1}\\
\hat{r}_{0}\\
-\hat{r}_{1}
\end{array}\right],\label{eq:Y1m}
\end{eqnarray}

\begin{eqnarray}
\left[\begin{array}{c}
Y_{22}\\
Y_{21}\\
Y_{20}\\
Y_{2-1}\\
Y_{2-2}
\end{array}\right] & = & \frac{1}{2}\sqrt{\frac{15}{2\pi}}\left[\sqrt{\frac{2}{3}}\begin{array}{c}
\hat{r}_{-1}^{2}\\
-\sqrt{2}\hat{r}_{0}\hat{r}_{-1}\\
\left(\hat{r}_{0}^{2}+\hat{r}_{-1}\hat{r}_{1}\right)\\
-\sqrt{2}\hat{r}_{0}\hat{r}_{1}\\
\hat{r}_{1}^{2}
\end{array}\right],\label{eq:Y2m}
\end{eqnarray}

\begin{eqnarray}
\left[\begin{array}{c}
Y_{33}\\
Y_{32}\\
Y_{31}\\
Y_{30}\\
Y_{3-1}\\
Y_{3-2}\\
Y_{3-3}
\end{array}\right] & = & \frac{1}{2}\sqrt{\frac{35}{2\pi}}\left[\begin{array}{c}
-\hat{r}_{-1}^{3}\\
\sqrt{3}\hat{r}_{0}\hat{r}_{-1}^{2}\\
-\sqrt{\frac{3}{5}}\left(2\hat{r}_{0}^{2}+\hat{r}_{1}\hat{r}_{-1}\right)\hat{r}_{-1}\\
\sqrt{\frac{2}{5}}\left(\hat{r}_{0}^{2}+3\hat{r}_{0}\hat{r}_{-1}\right)\hat{r}_{0}\\
-\sqrt{\frac{3}{5}}\left(2\hat{r}_{0}^{2}+\hat{r}_{1}\hat{r}_{-1}\right)\hat{r}_{1}\\
\sqrt{3}\hat{r}_{0}\hat{r}_{1}^{2}\\
-\hat{r}_{1}^{3}
\end{array}\right].\label{eq:Y3m}
\end{eqnarray}

\section{Electric dipole moment }

In this subsection, we derive the exact electric dipole moment for
the spherical and Cartesian coordinates.

\subsection{Spherical coordinates}

The electric dipole moment in spherical coordinate can be found by
using Eq.~\ref{eq:MM} ($j=1$, i.e. $\overline{l}=j\pm1=0,2$) :

\begin{eqnarray}
\left[\begin{array}{c}
a_{11}^{\omega}\\
a_{10}^{\omega}\\
a_{1-1}^{\omega}
\end{array}\right] & = & \left[\begin{array}{c}
a_{11}^{\omega}\\
a_{10}^{\omega}\\
a_{1-1}^{\omega}
\end{array}\right]^{\bar{l}=0}+\left[\begin{array}{c}
a_{11}^{\omega}\\
a_{10}^{\omega}\\
a_{1-1}^{\omega}
\end{array}\right]^{\bar{l}=2},\label{eq:ED_SC}\\
 & = & -\frac{1}{\pi\sqrt{3}}\int d^{3}\mathbf{r}\mathbf{J}^{\omega}j_{0}\left(kr\right)-\frac{k^{2}}{2\pi\sqrt{3}}\int d^{3}\mathbf{r}\left[3\left(\mathbf{r}^{\dagger}\mathbf{J}_{\omega}\left(\mathbf{r}\right)\right)\mathbf{r}-r^{2}\mathbf{J}_{\omega}\left(\mathbf{r}\right)\right]\frac{j_{2}\left(kr\right)}{\left(kr\right)^{2}},\nonumber
\end{eqnarray}

the derivation for the above expression can be found in our previous
work~\cite{FerCor2015b}. In the next section, we introduce the exact
Cartesian electric dipole moment.

\subsection{Cartesian coordinates}

Each components of the electric dipole moment (Eq.~\ref{eq:ED_SC})
in spherical basis can be written as:

\begin{eqnarray}
a_{11}^{\omega} & = & -\frac{1}{\pi\sqrt{3}}\int d^{3}\mathbf{r}J_{1}^{\omega}j_{0}\left(kr\right)+\frac{k^{2}}{2\pi\sqrt{3}}\int d^{3}\mathbf{r}\left[3J_{-1}^{\omega}r_{1}^{2}-3J_{0}^{\omega}r_{0}r_{1}+J_{1}^{\omega}\left(r_{0}^{2}+r_{-1}r_{0}\right)\right]\frac{j_{2}\left(kr\right)}{\left(kr\right)^{2}},\nonumber \\
a_{10}^{\omega} & = & -\frac{1}{\pi\sqrt{3}}\int d^{3}\mathbf{r}J_{0}^{\omega}j_{0}\left(kr\right)+\frac{k^{2}}{2\pi\sqrt{3}}\int d^{3}\mathbf{r}\left[3J_{1}^{\omega}r_{-1}r_{0}+3J_{-1}^{\omega}r_{0}r_{1}-2J_{0}^{\omega}\left(r_{0}^{2}+r_{-1}r_{0}\right)\right]\frac{j_{2}\left(kr\right)}{\left(kr\right)^{2}},\nonumber \\
a_{1-1}^{\omega} & = & -\frac{1}{\pi\sqrt{3}}\int d^{3}\mathbf{r}J_{-1}^{\omega}j_{0}\left(kr\right)+\frac{k^{2}}{2\pi\sqrt{3}}\int d^{3}\mathbf{r}\left[J_{-1}^{\omega}\left(r_{0}^{2}+r_{-1}r_{0}\right)-3J_{0}^{\omega}r_{-1}r_{0}+3J_{1}^{\omega}r_{-1}^{2}\right]\frac{j_{2}\left(kr\right)}{\left(kr\right)^{2}},\label{eq:ED_S}
\end{eqnarray}

Now, the Cartesian electric dipole moment can be found by using the
following transformation between the spherical and Cartesian coordinates:

\begin{eqnarray}
p_{x}^{\omega} & = & C_{1}^{e}\frac{a_{1-1}^{\omega}-a_{11}^{\omega}}{\sqrt{2}},\nonumber \\
p_{y}^{\omega} & = & C_{1}^{e}\frac{a_{1-1}^{\omega}+a_{11}^{\omega}}{\sqrt{2}i},\nonumber \\
p_{z}^{\omega} & = & C_{1}^{e}a_{10}^{\omega},\label{eq:ED_S2C}
\end{eqnarray}

where $C_{1}^{e}=\frac{\sqrt{3}\pi}{i\omega}$, which is obtained
by comparing the electric field expressions in spherical \cite{Jackson:98}
and Cartesian coordinates, i.e. Eq.~\ref{eq:E_Field}. Finally, we
substitute Eqs\@.~\ref{eq:ED_S} in Eqs.~\ref{eq:ED_S2C} and by
using Eqs.~\ref{eq:r_C2S} and Eqs.~\ref{eq:J_C2S}, we get:

\begin{eqnarray}
p_{x}^{\omega} & = & -\frac{1}{i\omega}\left\{ \int d^{3}\mathbf{r}J_{x}^{\omega}j_{0}\left(kr\right)+\frac{k^{2}}{2}\int d^{3}\mathbf{r}\left[3\left(\mathbf{r}\cdot\mathbf{J}_{\omega}\right)x-r^{2}J_{x}^{\omega}\right]\frac{j_{2}\left(kr\right)}{\left(kr\right)^{2}}\right\} ,\nonumber \\
p_{y}^{\omega} & = & -\frac{1}{i\omega}\left\{ \int d^{3}\mathbf{r}J_{y}^{\omega}j_{0}\left(kr\right)+\frac{k^{2}}{2}\int d^{3}\mathbf{r}\left[3\left(\mathbf{r}\cdot\mathbf{J}_{\omega}\right)y-r^{2}J_{y}^{\omega}\right]\frac{j_{2}\left(kr\right)}{\left(kr\right)^{2}}\right\} ,\\
p_{z}^{\omega} & = & -\frac{1}{i\omega}\left\{ \int d^{3}\mathbf{r}J_{z}^{\omega}j_{0}\left(kr\right)+\frac{k^{2}}{2}\int d^{3}\mathbf{r}\left[3\left(\mathbf{r}\cdot\mathbf{J}_{\omega}\right)z-r^{2}J_{z}^{\omega}\right]\frac{j_{2}\left(kr\right)}{\left(kr\right)^{2}}\right\} ,\nonumber
\end{eqnarray}

which can be written in a short form:

\[
\boxed{p_{\alpha}^{\omega}=-\frac{1}{i\omega}\left\{ \int d^{3}\mathbf{r}J_{\alpha}^{\omega}j_{0}\left(kr\right)+\frac{k^{2}}{2}\int d^{3}\mathbf{r}\left[3\left(\mathbf{r}\cdot\mathbf{J}_{\omega}\right)r_{\alpha}-r^{2}J_{\alpha}^{\omega}\right]\frac{j_{2}\left(kr\right)}{\left(kr\right)^{2}}\right\} }
\]

where $\alpha=x,y,z$. Note that the above expression for the electric
dipole moment is valid for \textit{any} wavelength. This expression
is documented in Tab. 2 of the main manuscript.

\subsection{Long-wavelength approximation }

We now can make a \textit{long-wavelength approximation} by using
the small argument approximation to the spherical Bessel function,
i.e.

\begin{eqnarray}
j_{0}\left(kr\right) & \approx & 1-\frac{\left(kr\right)^{2}}{6},\nonumber \\
j_{2}\left(kr\right) & \approx & \frac{\left(kr\right)^{2}}{15},
\end{eqnarray}

and obtain the expression for the approximate electric dipole moments
that are only valid for sources sufficiently small in their spatial
extent with respect to the wavelength:

\begin{eqnarray}
p_{\alpha}^{\omega} & = & -\frac{1}{i\omega}\left\{ \int d^{3}\mathbf{r}J_{\alpha}^{\omega}j_{0}\left(kr\right)+\frac{k^{2}}{2}\int d^{3}\mathbf{r}\left[3\left(\mathbf{r}\cdot\mathbf{J}_{\omega}\right)r_{\alpha}-r^{2}J_{\alpha}^{\omega}\right]\frac{j_{2}\left(kr\right)}{\left(kr\right)^{2}}\right\} ,\nonumber \\
 & \approx & -\frac{1}{i\omega}\left\{ \int d^{3}\mathbf{r}J_{\alpha}^{\omega}+\frac{k^{2}}{10}\int d^{3}\mathbf{r}\left[\left(\mathbf{r}\cdot\mathbf{J}_{\omega}\right)r_{\alpha}-2r^{2}J_{\alpha}^{\omega}\right]\right\} .
\end{eqnarray}

This expression is documented in Tab. 1 of the main manuscript. It
is important to note that the first term, i.e.

\begin{equation}
-\frac{1}{i\omega}\int d^{3}\mathbf{r}J_{\alpha}^{\omega},
\end{equation}

can be found in \textit{any} electrodynamics textbook~\cite{Jackson:98}.
It is often considered as the \textit{electric dipole momen}t. However,
we shall always keep in mind that this expression is \textit{only}
valid for very small particles compared to wavelength. The second
term

\begin{equation}
-\frac{1}{i\omega}\frac{k^{2}}{10}\int d^{3}\mathbf{r}\left[\left(\mathbf{r}\cdot\mathbf{J}_{\omega}\right)r_{\alpha}-2r^{2}J_{\alpha}^{\omega}\right],
\end{equation}

is the so-called toroidal dipole moment. They have been \textit{incorrectly}
called the \textit{third} family of multipole moments. However, based
on our derivation, it is obvious that the first and second terms belong
to the electric dipole moment~(see Ref.~\cite{FerCor2015b,FerCor2015c}).

\section{Magnetic dipole moment }

In this subsection, we derive the exact magnetic dipole moment for
the spherical and Cartesian coordinates.

\subsection{Spherical coordinates}

The magnetic dipole moment in spherical coordinate can be found by
using Eq.~\ref{eq:MM}($j=1$, i.e. $\overline{l}=j=1$) :

\begin{eqnarray}
\left[\begin{array}{c}
b_{11}^{\omega}\\
b_{10}^{\omega}\\
b_{1-1}^{\omega}
\end{array}\right] & = & \left[\begin{array}{c}
b_{11}^{\omega}\\
b_{10}^{\omega}\\
b_{1-1}^{\omega}
\end{array}\right]^{\bar{l}=1},\label{eq:MD_SC}\\
 & = & -\frac{\sqrt{3}k}{2\pi}\int d^{3}\mathbf{r}\mathbf{r}\times\mathbf{J_{\omega}}\frac{j_{1}\left(kr\right)}{kr},\nonumber
\end{eqnarray}

In contrast to the electric dipole moment, it has only one term (i.e.
$\bar{l}=1$). The derivation for the above expression can be found
in our previous work~\cite{FerCor2015b}. In the next subsection,
we introduce the exact magnetic dipole moment in Cartesian coordinates.

\subsection{Cartesian coordinates}

Each components of the magnetic dipole moment (Eq.~\ref{eq:ED_SC})
can be written as:

\begin{eqnarray}
b_{11}^{\omega} & = & -\frac{\sqrt{3}k}{2\pi}\int d^{3}\mathbf{r}\left(\mathbf{r}\times\mathbf{J_{\omega}}\right)_{1}\frac{j_{1}\left(kr\right)}{kr},\nonumber \\
b_{10}^{\omega} & = & -\frac{\sqrt{3}k}{2\pi}\int d^{3}\mathbf{r}\left(\mathbf{r}\times\mathbf{J_{\omega}}\right)_{0}\frac{j_{1}\left(kr\right)}{kr},\nonumber \\
b_{1-1}^{\omega} & = & -\frac{\sqrt{3}k}{2\pi}\int d^{3}\mathbf{r}\left(\mathbf{r}\times\mathbf{J_{\omega}}\right)_{-1}\frac{j_{1}\left(kr\right)}{kr}.\label{eq:MD_S}
\end{eqnarray}

The Cartesian magnetic dipole moment can be obtained by using the
following transformation between the spherical and Cartesian coordinates:

\begin{eqnarray}
m_{x}^{\omega} & = & C_{1}^{m}\frac{b_{1-1}^{\omega}-b_{11}^{\omega}}{\sqrt{2}},\nonumber \\
m_{y}^{\omega} & = & C_{1}^{m}\frac{b_{1-1}^{\omega}+b_{11}^{\omega}}{\sqrt{2}i},\nonumber \\
m_{z}^{\omega} & = & C_{1}^{m}b_{10}^{\omega},\label{eq:ED_S2C-1}
\end{eqnarray}

where $C_{1}^{m}=-\frac{\sqrt{3}\pi}{k}$, which is obtained by comparing
the electric field expressions in spherical \cite{Jackson:98} and
Cartesian coordinates, i.e. Eq.~\ref{eq:E_Field}. Finally, we substitute
in Eqs\@.~\ref{eq:MD_S} in Eqs.~\ref{eq:ED_S2C-1} and by using
Eq.~\ref{eq:r_C2S} and Eq.~\ref{eq:J_C2S}, we get:

\begin{eqnarray}
m_{x}^{\omega} & = & \frac{3}{2}\int d^{3}\mathbf{r}\left(\mathbf{r}\times\mathbf{J_{\omega}}\right)_{x}\frac{j_{1}\left(kr\right)}{kr},\nonumber \\
m_{y}^{\omega} & = & \frac{3}{2}\int d^{3}\mathbf{r}\left(\mathbf{r}\times\mathbf{J_{\omega}}\right)_{y}\frac{j_{1}\left(kr\right)}{kr},\nonumber \\
m_{z}^{\omega} & = & \frac{3}{2}\int d^{3}\mathbf{r}\left(\mathbf{r}\times\mathbf{J_{\omega}}\right)_{z}\frac{j_{1}\left(kr\right)}{kr},
\end{eqnarray}

which can be written in a short form:

\begin{equation}
\boxed{m_{\alpha}^{\omega}=\frac{3}{2}\int d^{3}\mathbf{r}\left(\mathbf{r}\times\mathbf{J_{\omega}}\right)_{\alpha}\frac{j_{1}\left(kr\right)}{kr}}\label{eq:MD_exact}
\end{equation}

where $\alpha=x,y,z$. This expression is documented in Tab. 2 of
the main manuscript.

\subsection{Long-wavelength approximation }

We now can make the \textit{long-wavelength approximation} by using
the small argument approximation to the spherical Bessel function,
i.e.

\begin{eqnarray}
j_{1}\left(kr\right) & \approx & \frac{kr}{3},
\end{eqnarray}

and obtain the well-known long-wavelength approximation expression
for the magnetic dipole moments:

\begin{eqnarray}
m_{\alpha}^{\omega} & = & \frac{3}{2}\int d^{3}\mathbf{r}\left(\mathbf{r}\times\mathbf{J_{\omega}}\right)_{\alpha}\frac{j_{1}\left(kr\right)}{kr},\nonumber \\
 & \approx & \frac{1}{2}\int d^{3}\mathbf{r}\left(\mathbf{r}\times\mathbf{J_{\omega}}\right)_{\alpha}.
\end{eqnarray}

This expression is documented in Tab. 1 of the main manuscript. It
is important to mention that the long-wavelength expression can be
found in \textit{any} electrodynamics textbook~\cite{Jackson:98}
and \textit{only} valid at the small object compared to wavelength
(i.e. $kr\ll1$), whereas our new expression is valid irrespective
of the size of the source.

\section{Magnetic quadrupole moment}

In this section, we derive the exact magnetic quadrupole moment for
the spherical and Cartesian coordinates by using Eq\@.~\ref{eq:MM}.
They have not been reported neither in spherical nor in Cartesian
coordinates

\subsection{Spherical coordinates}

The magnetic quadrupole moment in spherical coordinate can be found
by using Eq.~\ref{eq:EM} ($j=2$, i.e. $\overline{l}=j=2$), i.e.

\begin{eqnarray}
b_{2m}^{\omega} & = & -\frac{4\pi}{\sqrt{\left(2\pi\right)^{3}}}\stackrel[\overline{m}=-2]{2}{\sum}\int d\widehat{\mathbf{p}}\mathbf{X}_{2m}^{\dagger}\left(\widehat{\mathbf{p}}\right)Y_{2\overline{m}}\left(\widehat{\mathbf{p}}\right)\int d^{3}\mathbf{r}\mathbf{J_{\omega}\left(r\right)}Y_{2\overline{m}}^{*}\left(\widehat{\mathbf{r}}\right)j_{2}\left(kr\right).
\end{eqnarray}

By using the explicit expression for the multipolar functions in momentum
space, i.e.

\begin{eqnarray}
\mathbf{X}_{jm}^{\dagger}\left(\widehat{\mathbf{p}}\right) & = & \frac{1}{\sqrt{j\left(j+1\right)}}\left[\begin{array}{c}
-\sqrt{\frac{j\left(j+1\right)-m\left(m-1\right)}{2}}Y_{j\left(m-1\right)}\left(\widehat{\mathbf{p}}\right)\\
mY_{jm}\left(\widehat{\mathbf{p}}\right)\\
\sqrt{\frac{j\left(j+1\right)-m\left(m+1\right)}{2}}Y_{j\left(m+1\right)}\left(\widehat{\mathbf{p}}\right)
\end{array}\right],\label{eq:X_Def}
\end{eqnarray}

we can compute the momentum integrals ($\int d\widehat{\mathbf{p}}\mathbf{X}_{2m}^{\dagger}\left(\widehat{\mathbf{p}}\right)Y_{2\overline{m}}\left(\widehat{\mathbf{p}}\right)$)
for each $m$ case
\begin{equation}
m=2\rightarrow\int d\widehat{\mathbf{p}}\mathbf{X}_{22}^{\dagger}\left(\widehat{\mathbf{p}}\right)Y_{2\bar{m}}\left(\widehat{\mathbf{p}}\right)\Rightarrow\begin{array}{cc}
\bar{m}=2 & \rightarrow\\
\bar{m}=1 & \rightarrow\\
\bar{m}=0 & \rightarrow\\
\bar{m}=-1 & \rightarrow\\
\bar{m}=-2 & \rightarrow
\end{array}\mbox{\ensuremath{\frac{1}{\sqrt{3}}}}\begin{array}{ccccc}
( & 0 & \sqrt{2} & 0 & )\\
( & -1 & 0 & 0 & )\\
( & 0 & 0 & 0 & ),\\
( & 0 & 0 & 0 & )\\
( & 0 & 0 & 0 & )
\end{array}\label{eq:M_m_2}
\end{equation}

\begin{equation}
m=1\rightarrow\int d\widehat{\mathbf{p}}\mathbf{X}_{21}^{\dagger}\left(\widehat{\mathbf{p}}\right)Y_{2\bar{m}}\left(\widehat{\mathbf{p}}\right)\Rightarrow\begin{array}{cc}
\bar{m}=2 & \rightarrow\\
\bar{m}=1 & \rightarrow\\
\bar{m}=0 & \rightarrow\\
\bar{m}=-1 & \rightarrow\\
\bar{m}=-2 & \rightarrow
\end{array}\mbox{\ensuremath{\frac{1}{\sqrt{6}}}}\begin{array}{ccccc}
( & 0 & 0 & \sqrt{2} & )\\
( & 0 & 1 & 0 & )\\
( & -\sqrt{3} & 0 & 0 & ),\\
( & 0 & 0 & 0 & )\\
( & 0 & 0 & 0 & )
\end{array}
\end{equation}

\begin{equation}
m=0\rightarrow\int d\widehat{\mathbf{p}}\mathbf{X}_{20}^{\dagger}\left(\widehat{\mathbf{p}}\right)Y_{2\bar{m}}\left(\widehat{\mathbf{p}}\right)\Rightarrow\begin{array}{cc}
\bar{m}=2 & \rightarrow\\
\bar{m}=1 & \rightarrow\\
\bar{m}=0 & \rightarrow\\
\bar{m}=-1 & \rightarrow\\
\bar{m}=-2 & \rightarrow
\end{array}\mbox{\ensuremath{\frac{1}{\sqrt{2}}}}\begin{array}{ccccc}
( & 0 & 0 & 0 & )\\
( & 0 & 0 & 1 & )\\
( & 0 & 0 & 0 & ),\\
( & -1 & 0 & 0 & )\\
( & 0 & 0 & 0 & )
\end{array}
\end{equation}

\begin{equation}
m=-1\rightarrow\int d\widehat{\mathbf{p}}\mathbf{X}_{2-1}^{\dagger}\left(\widehat{\mathbf{p}}\right)Y_{2\bar{m}}\left(\widehat{\mathbf{p}}\right)\Rightarrow\begin{array}{cc}
\bar{m}=2 & \rightarrow\\
\bar{m}=1 & \rightarrow\\
\bar{m}=0 & \rightarrow\\
\bar{m}=-1 & \rightarrow\\
\bar{m}=-2 & \rightarrow
\end{array}\mbox{\ensuremath{\frac{-1}{\sqrt{6}}}}\begin{array}{ccccc}
( & 0 & 0 & 0 & )\\
( & 0 & 0 & 0 & )\\
( & 0 & 0 & -\sqrt{3} & )\\
( & 0 & 1 & 0 & )\\
( & \sqrt{2} & 0 & 0 & )
\end{array},
\end{equation}

\begin{equation}
m=-2\rightarrow\int d\widehat{\mathbf{p}}\mathbf{X}_{2-2}^{\dagger}\left(\widehat{\mathbf{p}}\right)Y_{2\bar{m}}\left(\widehat{\mathbf{p}}\right)\Rightarrow\begin{array}{cc}
\bar{m}=2 & \rightarrow\\
\bar{m}=1 & \rightarrow\\
\bar{m}=0 & \rightarrow\\
\bar{m}=-1 & \rightarrow\\
\bar{m}=-2 & \rightarrow
\end{array}\mbox{\ensuremath{\frac{-1}{\sqrt{3}}}}\begin{array}{ccccc}
( & 0 & 0 & 0 & )\\
( & 0 & 0 & 0 & )\\
( & 0 & 0 & 0 & )\\
( & 0 & 0 & -1 & )\\
( & 0 & \sqrt{2} & 0 & )
\end{array}.\label{eq:M_m2}
\end{equation}

Note that we list row vectors for each $\bar{m}=-2,-1,0,1,2$. The
summation in $\bar{m}$ in Eq.~\ref{eq:M_m_2}-\ref{eq:M_m2} can
be done and by using Eq.~\ref{eq:J_and_r_def} and Eq.~\ref{eq:Y2m},
after some simple algebra, we get:

\begin{eqnarray}
b_{22}^{\omega} & = & \sqrt{\frac{5}{2\pi^{2}}}\int d^{3}\mathbf{r}\hat{r}_{1}\left(J_{1}^{\omega}\hat{r}_{0}-J_{0}^{\omega}\hat{r}_{1}\right)j_{2}\left(kr\right),\nonumber \\
b_{21}^{\omega} & = & \sqrt{\frac{5}{4\pi^{2}}}\int d^{3}\mathbf{r}\left[\hat{r}_{1}\left(J_{1}^{\omega}\hat{r}_{-1}-J_{-1}^{\omega}\hat{r}_{1}\right)+\hat{r}_{0}\left(J_{1}^{\omega}\hat{r}_{0}-J_{0}^{\omega}\hat{r}_{1}\right)\right]j_{2}\left(kr\right),\nonumber \\
b_{20}^{\omega} & = & \sqrt{\frac{15}{4\pi^{2}}}\int d^{3}\mathbf{r}\hat{r}_{0}\left(J_{1}^{\omega}\hat{r}_{-1}-J_{-1}^{\omega}\hat{r}_{1}\right)j_{2}\left(kr\right),\nonumber \\
b_{2-1}^{\omega} & = & \sqrt{\frac{5}{4\pi^{2}}}\int d^{3}\mathbf{r}\left[\hat{r}_{-1}\left(J_{1}^{\omega}\hat{r}_{-1}-J_{-1}^{\omega}\hat{r}_{1}\right)+\hat{r}_{0}\left(J_{0}^{\omega}\hat{r}_{-1}-J_{-1}^{\omega}\hat{r}_{0}\right)\right]j_{2}\left(kr\right),\nonumber \\
b_{2-2}^{\omega} & = & \sqrt{\frac{5}{2\pi^{2}}}\int d^{3}\mathbf{r}\hat{r}_{-1}\left(J_{0}^{\omega}\hat{r}_{-1}-J_{-1}^{\omega}\hat{r}_{0}\right)j_{2}\left(kr\right).
\end{eqnarray}

Finally, by using the cross product definition in spherical basis,
i.e. Eq.~\ref{eq:CP_S}, the magnetic quadrupole moments in spherical
coordinates can be written as

\begin{eqnarray}
b_{22}^{\omega} & = & \frac{5ik^{2}}{\pi\sqrt{10}}\int d^{3}\mathbf{r}\left[r_{1}\left(\mathbf{r}\times\mathbf{J_{\omega}}\right)_{1}\right]\frac{j_{2}\left(kr\right)}{\left(kr\right)^{2}},\nonumber \\
b_{21}^{\omega} & = & \frac{5ik^{2}}{2\pi\sqrt{5}}\int d^{3}\mathbf{r}\left[r_{1}\left(\mathbf{r}\times\mathbf{J_{\omega}}\right)_{0}+\hat{r}_{0}\left(\widehat{\mathbf{r}}\times\mathbf{J_{\omega}}\right)_{1}\right]\frac{j_{2}\left(kr\right)}{\left(kr\right)^{2}},\nonumber \\
b_{20}^{\omega} & = & \frac{\sqrt{15}ik^{2}}{2\pi}\int d\mathbf{^{3}r}\left[r_{0}\left(\mathbf{r}\times\mathbf{J_{\omega}}\right)_{0}\right]\frac{j_{2}\left(kr\right)}{\left(kr\right)^{2}},\nonumber \\
b_{2-1}^{\omega} & = & \frac{5ik^{2}}{2\pi\sqrt{5}}\int d^{3}\mathbf{r}\left[r_{-1}\left(\mathbf{r}\times\mathbf{J_{\omega}}\right)_{0}+r_{0}\left(\widehat{\mathbf{r}}\times\mathbf{J_{\omega}}\right)_{-1}\right]\frac{j_{2}\left(kr\right)}{\left(kr\right)^{2}},\nonumber \\
b_{2-2}^{\omega} & = & \frac{5ik^{2}}{\pi\sqrt{10}}\int d^{3}\mathbf{r}\left[r_{-1}\left(\mathbf{r}\times\mathbf{J_{\omega}}\right)_{-1}\right]\frac{j_{2}\left(kr\right)}{\left(kr\right)^{2}}.\label{eq:MQ_S}
\end{eqnarray}

\subsection{Cartesian coordinates}

In order to obtain the magnetic quadrupole moments in the Cartesian
coordinates, we use the following transformation between the spherical
and Cartesian coordinates:

\begin{eqnarray}
Q_{xx}^{m} & = & C_{2}^{\mathrm{m}}\left[\frac{b_{22}^{\omega}+b_{2-2}^{\omega}}{2}-\frac{1}{\sqrt{6}}b_{20}^{\omega}\right],\nonumber \\
Q_{xy}^{m}=Q_{yx}^{m} & = & C_{2}^{\mathrm{m}}\left[\frac{b_{2-2}^{\omega}-b_{22}^{\omega}}{2i}\right],\nonumber \\
Q_{xz}^{m}=Q_{zx}^{m} & = & C_{2}^{\mathrm{m}}\left[\frac{b_{2-1}^{\omega}-b_{21}^{\omega}}{2}\right],\label{eq:T_EQ}\\
Q_{yz}^{m}=Q_{zy}^{m} & = & C_{2}^{\mathrm{m}}\left[\frac{b_{2-1}^{\omega}+b_{21}^{\omega}}{2i}\right],\nonumber \\
Q_{yy}^{m} & = & C_{2}^{\mathrm{m}}\left[-\left(\frac{b_{22}^{\omega}+b_{2-2}^{\omega}}{2}\right)-\frac{1}{\sqrt{6}}b_{20}^{\omega}\right],\nonumber \\
Q_{zz}^{m} & = & C_{2}^{\mathrm{m}}\left[\frac{2}{\sqrt{6}}b_{20}^{\omega}\right]=-Q_{xx}^{m}-Q_{yy}^{m},\nonumber
\end{eqnarray}

where $C_{2}^{\mathrm{m}}=\frac{6\pi\sqrt{10}}{ik^{2}}$, which is
obtained by comparing the electric field expressions in spherical
\cite{Jackson:98} and Cartesian coordinates, i.e. Eq.~\ref{eq:E_Field}.
It is important to note that $\mathbf{Q}^{m}$ is a symmetric tensor
and traceless, i.e. $Q_{xx}^{m}+Q_{yy}^{m}+Q_{zz}^{m}=0$, which reduces
the Cartesian quadrupolar moments to five independent components.

Now, by substitute Eq.~\ref{eq:MQ_S} in Eq.~\ref{eq:T_EQ}, and
by using Eq.~\ref{eq:CP_S} and Eq.~\ref{eq:r_C2S}. The final results
read as

\begin{eqnarray}
Q_{xx}^{m} & = & C_{2}^{\mathrm{m}}\left[\frac{b_{22}^{\omega}+b_{2-2}^{\omega}}{2}-\frac{1}{\sqrt{6}}b_{20}^{\omega}\right],\nonumber \\
 & = & C_{2}^{\mathrm{m}}\frac{5ik^{2}}{2\pi\sqrt{10}}\int d^{3}\mathbf{r}\left[r_{1}\left(\mathbf{r}\times\mathbf{J_{\omega}}\right)_{1}-r_{-1}\left(\mathbf{r}\times\mathbf{J_{\omega}}\right)_{-1}-r_{0}\left(\mathbf{r}\times\mathbf{J_{\omega}}\right)_{0}\right]\frac{j_{2}\left(kr\right)}{\left(kr\right)^{2}},\nonumber \\
 & = & 15\int d^{3}\mathbf{r}\left\{ \left(r_{-1}-r_{1}\right)\left[\left(\mathbf{r}\times\mathbf{J_{\omega}}\right)_{-1}-\left(\mathbf{r}\times\mathbf{J_{\omega}}\right)_{1}\right]\right\} \frac{j_{2}\left(kr\right)}{\left(kr\right)^{2}},\nonumber \\
 & = & 15\int d^{3}\mathbf{r}\left\{ x\left(\mathbf{r}\times\mathbf{J_{\omega}}\right)_{x}+x\left(\mathbf{r}\times\mathbf{J_{\omega}}\right)_{x}\right\} \frac{j_{2}\left(kr\right)}{\left(kr\right)^{2}},
\end{eqnarray}

\begin{eqnarray}
Q_{xy}^{m} & = & C_{2}^{\mathrm{m}}\left[\frac{b_{2-2}^{\omega}-b_{22}^{\omega}}{2i}\right],\nonumber \\
 & = & C_{2}^{\mathrm{m}}\frac{5k^{2}}{2\pi\sqrt{10}}\int d^{3}\mathbf{r}\left[r_{-1}\left(\mathbf{r}\times\mathbf{J_{\omega}}\right)_{-1}-r_{1}\left(\mathbf{r}\times\mathbf{J_{\omega}}\right)_{1}\right]\frac{j_{2}\left(kr\right)}{\left(kr\right)^{2}},\nonumber \\
 & = & 15\int d^{3}\mathbf{r}\left\{ x\left[\frac{\left(\mathbf{r}\times\mathbf{J_{\omega}}\right)_{-1}+\left(\mathbf{r}\times\mathbf{J_{\omega}}\right)_{1}}{\sqrt{2}i}\right]+y\left[\frac{\left(\mathbf{r}\times\mathbf{J_{\omega}}\right)_{-1}-\left(\mathbf{r}\times\mathbf{J_{\omega}}\right)_{1}}{\sqrt{2}}\right]y\right\} \frac{j_{2}\left(kr\right)}{\left(kr\right)^{2}},\nonumber \\
 & = & 15\int d^{3}\mathbf{r}\left\{ x\left(\mathbf{r}\times\mathbf{J_{\omega}}\right)_{y}+y\left(\mathbf{r}\times\mathbf{J_{\omega}}\right)_{x}\right\} \frac{j_{2}\left(kr\right)}{\left(kr\right)^{2}},
\end{eqnarray}

\begin{eqnarray}
Q_{xz}^{m} & = & C_{2}^{\mathrm{m}}\left[\frac{b_{2-1}^{\omega}-b_{21}^{\omega}}{2}\right],\nonumber \\
 & = & C_{2}^{\mathrm{m}}\frac{5ik^{2}}{2\pi\sqrt{2}\sqrt{10}}\int d^{3}\mathbf{r}\left\{ r_{-1}\left(\mathbf{r}\times\mathbf{J_{\omega}}\right)_{0}+r_{0}\left(\mathbf{r}\times\mathbf{J_{\omega}}\right)_{-1}-r_{1}\left(\mathbf{r}\times\mathbf{J_{\omega}}\right)_{0}-r_{0}\left(\mathbf{r}\times\mathbf{J_{\omega}}\right)_{1}\right\} \frac{j_{2}\left(kr\right)}{\left(kr\right)^{2}},\nonumber \\
 & = & \frac{15}{\sqrt{2}}\int d^{3}\mathbf{r}\left\{ \left(r_{-1}-r_{1}\right)\left(\mathbf{r}\times\mathbf{J_{\omega}}\right)_{0}+r_{0}\left[\left(\mathbf{r}\times\mathbf{J_{\omega}}\right)_{1}-\left(\mathbf{r}\times\mathbf{J_{\omega}}\right)_{-1}\right]\right\} \frac{j_{2}\left(kr\right)}{\left(kr\right)^{2}},\nonumber \\
 & = & 15\int d^{3}\mathbf{r}\left\{ \frac{\left(r_{-1}-r_{1}\right)}{\sqrt{2}}\left(\mathbf{r}\times\mathbf{J_{\omega}}\right)_{0}+r_{0}\frac{\left[\left(\mathbf{r}\times\mathbf{J_{\omega}}\right)_{-1}-\left(\mathbf{r}\times\mathbf{J_{\omega}}\right)_{1}\right]}{\sqrt{2}}\right\} \frac{j_{2}\left(kr\right)}{\left(kr\right)^{2}},\nonumber \\
 & = & 15\int d^{3}\mathbf{r}\left\{ x\left(\mathbf{r}\times\mathbf{J_{\omega}}\right)_{z}+z\left(\mathbf{r}\times\mathbf{J_{\omega}}\right)_{x}\right\} \frac{j_{2}\left(kr\right)}{\left(kr\right)^{2}},
\end{eqnarray}

\begin{eqnarray}
Q_{yy}^{m} & = & C_{2}^{\mathrm{m}}\left[\frac{b_{22}^{\omega}-b_{2-2}^{\omega}}{2}-\frac{1}{\sqrt{6}}b_{20}^{\omega}\right],\nonumber \\
 & = & C_{2}^{\mathrm{m}}\frac{5k^{2}}{2\pi\sqrt{10}}\int d^{3}\mathbf{r}\left[r_{-1}\left(\mathbf{r}\times\mathbf{J_{\omega}}\right)_{-1}-r_{1}\left(\mathbf{r}\times\mathbf{J_{\omega}}\right)_{1}-r_{0}\left(\mathbf{r}\times\mathbf{J_{\omega}}\right)_{0}\right]\frac{j_{2}\left(kr\right)}{\left(kr\right)^{2}},\nonumber \\
 & = & \frac{15}{i}\int d^{3}\mathbf{r}\left\{ -\left(r_{-1}+r_{1}\right)\left[\left(\mathbf{r}\times\mathbf{J_{\omega}}\right)_{-1}+\left(\mathbf{r}\times\mathbf{J_{\omega}}\right)_{1}\right]\right\} \frac{j_{2}\left(kr\right)}{\left(kr\right)^{2}},\nonumber \\
 & = & 15\int d^{3}\mathbf{r}\left\{ 2\frac{\left(r_{-1}+r_{1}\right)}{\sqrt{2}i}\left[\frac{\left(\mathbf{r}\times\mathbf{J_{\omega}}\right)_{-1}+\left(\mathbf{r}\times\mathbf{J_{\omega}}\right)_{1}}{\sqrt{2}i}\right]\right\} \frac{j_{2}\left(kr\right)}{\left(kr\right)^{2}},\nonumber \\
 & = & 15\int d^{3}\mathbf{r}\left\{ y\left(\mathbf{r}\times\mathbf{J_{\omega}}\right)_{y}+y\left(\mathbf{r}\times\mathbf{J_{\omega}}\right)_{y}\right\} \frac{j_{2}\left(kr\right)}{\left(kr\right)^{2}},
\end{eqnarray}

\begin{eqnarray}
Q_{yz}^{m} & = & C_{2}^{\mathrm{m}}\left[\frac{b_{2-1}^{\omega}+b_{21}^{\omega}}{2i}\right],\nonumber \\
 & = & C_{2}^{\mathrm{m}}\frac{5k^{2}}{2\pi\sqrt{2}\sqrt{10}}\int d^{3}\mathbf{r}\left\{ r_{-1}\left(\mathbf{r}\times\mathbf{J_{\omega}}\right)_{0}+r_{0}\left(\mathbf{r}\times\mathbf{J_{\omega}}\right)_{-1}+r_{1}\left(\mathbf{r}\times\mathbf{J_{\omega}}\right)_{0}+r_{0}\left(\mathbf{r}\times\mathbf{J_{\omega}}\right)_{1}\right\} \frac{j_{2}\left(kr\right)}{\left(kr\right)^{2}},\nonumber \\
 & = & \frac{15}{\sqrt{2}i}\int d^{3}\mathbf{r}\left\{ \left(r_{-1}+r_{1}\right)\left(\mathbf{r}\times\mathbf{J_{\omega}}\right)_{0}+r_{0}\left[\left(\mathbf{r}\times\mathbf{J_{\omega}}\right)_{1}+\left(\mathbf{r}\times\mathbf{J_{\omega}}\right)_{-1}\right]\right\} \frac{j_{2}\left(kr\right)}{\left(kr\right)^{2}},\nonumber \\
 & = & 15\int d^{3}\mathbf{r}\left\{ \left(\frac{r_{-1}+r_{1}}{\sqrt{2}i}\right)\left(\mathbf{r}\times\mathbf{J_{\omega}}\right)_{0}+r_{0}\left[\frac{\left(\mathbf{r}\times\mathbf{J_{\omega}}\right)_{1}+\left(\mathbf{r}\times\mathbf{J_{\omega}}\right)_{-1}}{\sqrt{2}i}\right]\right\} \frac{j_{2}\left(kr\right)}{\left(kr\right)^{2}},\nonumber \\
 & = & 15\int d^{3}\mathbf{r}\left\{ y\left(\mathbf{r}\times\mathbf{J_{\omega}}\right)_{z}+z\left(\mathbf{r}\times\mathbf{J_{\omega}}\right)_{x}\right\} \frac{j_{2}\left(kr\right)}{\left(kr\right)^{2}},
\end{eqnarray}

\begin{eqnarray}
Q_{zz}^{m} & = & C_{2}^{\mathrm{m}}\left[\frac{2}{\sqrt{6}}b_{20}^{\omega}\right],\nonumber \\
 & = & C_{2}^{\mathrm{m}}\frac{5ik^{2}}{2\pi\sqrt{10}}\int d^{3}\mathbf{r}\left[r_{0}\left(\mathbf{r}\times\mathbf{J_{\omega}}\right)_{0}+r_{0}\left(\mathbf{r}\times\mathbf{J_{\omega}}\right)_{0}\right]\frac{j_{2}\left(kr\right)}{\left(kr\right)^{2}},\nonumber \\
 & = & 15\int d^{3}\mathbf{r}\left\{ z\left(\mathbf{r}\times\mathbf{J_{\omega}}\right)_{z}+z\left(\mathbf{r}\times\mathbf{J_{\omega}}\right)_{z}\right\} \frac{j_{2}\left(kr\right)}{\left(kr\right)^{2}}.
\end{eqnarray}

Finally, we can write the above expressions for the Cartesian magnetic
quadrupole moment in a short form:

\[
\boxed{Q_{\alpha\beta}^{m}=15\int d^{3}\mathbf{r}\left\{ r_{\alpha}\left(\mathbf{r}\times\mathbf{J_{\omega}}\right)_{\beta}+r_{\beta}\left(\mathbf{r}\times\mathbf{J_{\omega}}\right)_{\alpha}\right\} \frac{j_{2}\left(kr\right)}{\left(kr\right)^{2}}}
\]

where $\alpha,\beta=x,y,z$. This expression is documented in Tab.
2 of the main manuscript.

\subsection{Long-wavelength approximation }

We now can make the \textit{long-wavelength approximation} by using
the small argument approximation to the spherical Bessel function,
i.e.

\begin{eqnarray}
j_{2}\left(kr\right) & \approx & \left(kr\right)^{2}/15,
\end{eqnarray}

and obtain the well-known long-wavelength expression for the magnetic
quadrupole moments:

\begin{eqnarray}
Q_{\alpha\beta}^{\mathrm{m}} & = & 15\int d^{3}\mathbf{r}\left\{ r_{\alpha}\left(\mathbf{r}\times\mathbf{J_{\omega}}\right)_{\beta}+r_{\beta}\left(\mathbf{r}\times\mathbf{J_{\omega}}\right)_{\alpha}\right\} \frac{j_{2}\left(kr\right)}{\left(kr\right)^{2}},\nonumber \\
 & \approx & \int d^{3}\mathbf{r}\left\{ r_{\alpha}\left(\mathbf{r}\times\mathbf{J_{\omega}}\right)_{\beta}+r_{\beta}\left(\mathbf{r}\times\mathbf{J_{\omega}}\right)_{\alpha}\right\} .
\end{eqnarray}

This expression is documented in Tab. 1 of the main manuscript. It
is important to mention that this expression can be found in \textit{any}
electrodynamics textbook~\cite{Jackson:98} and \textit{only} valid
at the long-wavelength regime.

\section{Electric quadrupole moment}

In this section, we derive the exact electric quadrupole moment in
the spherical and Cartesian coordinates by using Eq\@.~\ref{eq:EM}.

\subsection{Spherical coordinates}

The electric quadrupole moment in spherical coordinate has two contributions
($j=2$, i.e. $\overline{l}=j\pm1=1,3$) , i.e.

\begin{eqnarray}
a_{2m}^{\omega} & = & a_{2m}^{\omega\,\,\bar{l}=1}+a_{2m}^{\omega\,\,\bar{l}=3},\label{eq:ED_SC-1}\\
\left[\begin{array}{c}
a_{22}^{\omega}\\
a_{21}^{\omega}\\
a_{20}^{\omega}\\
a_{2-1}^{\omega}\\
a_{2-2}^{\omega}
\end{array}\right] & = & \left[\begin{array}{c}
a_{22}^{\omega}\\
a_{21}^{\omega}\\
a_{20}^{\omega}\\
a_{2-1}^{\omega}\\
a_{2-2}^{\omega}
\end{array}\right]^{\bar{l}=1}+\left[\begin{array}{c}
a_{22}^{\omega}\\
a_{21}^{\omega}\\
a_{20}^{\omega}\\
a_{2-1}^{\omega}\\
a_{2-2}^{\omega}
\end{array}\right]^{\bar{l}=3},
\end{eqnarray}

\subsubsection{$a_{2m}^{\omega\,\,\bar{l}=1}$ contribution}

In order to derive the $\,a_{2m}^{\omega\,\,\bar{l}=1}$ contribution
of the electric quadrupole moment in spherical coordinate, we use
Eq.~\ref{eq:EM}, i.e.

\begin{eqnarray*}
a_{2m}^{\omega\,\,\bar{l}=1} & = & -\frac{4\pi i}{\sqrt{\left(2\pi\right)^{3}}}\stackrel[\overline{m}=-1]{1}{\sum}\int d\widehat{\mathbf{p}}\mathbf{Z}_{2m}^{\dagger}\left(\widehat{\mathbf{p}}\right)Y_{1\overline{m}}\left(\widehat{\mathbf{p}}\right)\int d^{3}\mathbf{r}\mathbf{J_{\omega}\left(r\right)}Y_{1\overline{m}}^{*}\left(\widehat{\mathbf{r}}\right)j_{1}\left(kr\right),
\end{eqnarray*}

by the using the spherical function $Y_{2\overline{m}}$ and multipolar
functions in momentum space, i.e. $\mathbf{Z}_{jm}\left(\widehat{\mathbf{p}}\right)$

\begin{eqnarray}
\mathbf{Z}_{jm}\left(\widehat{\mathbf{p}}\right) & = & i\widehat{\mathbf{p}}\times\mathbf{X}_{jm}\left(\widehat{\mathbf{p}}\right),\label{eq:Zjm}\\
\hat{\mathbf{p}} & = & \frac{\mathbf{p}}{\left|\mathbf{p}\right|}=2\sqrt{\frac{\pi}{3}}\left[\begin{array}{c}
-Y_{1-1}\\
Y_{10}\\
-Y_{11}
\end{array}\right],
\end{eqnarray}

and having the explicit expression for $\mathbf{X}_{jm}\left(\widehat{\mathbf{p}}\right)$
in Eq.~\ref{eq:X_Def}, we can compute the momentum integrals {[}$\int d\widehat{\mathbf{p}}\mathbf{Z}_{2m}^{\dagger}\left(\widehat{\mathbf{p}}\right)Y_{1\bar{m}}\left(\widehat{\mathbf{p}}\right)${]}
for each $m$ case

\begin{equation}
m=2\rightarrow\int d\widehat{\mathbf{p}}\mathbf{Z}_{22}^{\dagger}\left(\widehat{\mathbf{p}}\right)Y_{1\bar{m}}\left(\widehat{\mathbf{p}}\right)\Rightarrow\begin{array}{cc}
\bar{m}=1 & \rightarrow\\
\bar{m}=0 & \rightarrow\\
\bar{m}=-1 & \rightarrow
\end{array}\mbox{\ensuremath{\frac{-1}{\sqrt{5}}}}\begin{array}{ccccc}
( & \sqrt{3} & 0 & 0 & )\\
( & 0 & 0 & 0 & )\\
( & 0 & 0 & 0 & )
\end{array},\label{eq:E_m_2}
\end{equation}

\begin{equation}
m=1\rightarrow\int d\widehat{\mathbf{p}}\mathbf{Z}_{21}^{\dagger}\left(\widehat{\mathbf{p}}\right)Y_{1\bar{m}}\left(\widehat{\mathbf{p}}\right)\Rightarrow\begin{array}{cc}
\bar{m}=1 & \rightarrow\\
\bar{m}=0 & \rightarrow\\
\bar{m}=-1 & \rightarrow
\end{array}\mbox{\ensuremath{\frac{-1}{\sqrt{10}}}}\begin{array}{ccccc}
( & 0 & \sqrt{3} & 0 & )\\
( & \sqrt{3} & 0 & 0 & )\\
( & 0 & 0 & 0 & )
\end{array},
\end{equation}

\begin{equation}
m=0\rightarrow\int d\widehat{\mathbf{p}}\mathbf{Z}_{20}^{\dagger}\left(\widehat{\mathbf{p}}\right)Y_{1\bar{m}}\left(\widehat{\mathbf{p}}\right)\Rightarrow\begin{array}{cc}
\bar{m}=1 & \rightarrow\\
\bar{m}=0 & \rightarrow\\
\bar{m}=-1 & \rightarrow
\end{array}\mbox{\ensuremath{\frac{-1}{\sqrt{10}}}}\begin{array}{ccccc}
( & 0 & 0 & 1 & )\\
( & 0 & 2 & 0 & )\\
( & 1 & 0 & 0 & )
\end{array},
\end{equation}

\begin{equation}
m=-1\rightarrow\int d\widehat{\mathbf{p}}\mathbf{Z}_{2-1}^{\dagger}\left(\widehat{\mathbf{p}}\right)Y_{1\bar{m}}\left(\widehat{\mathbf{p}}\right)\Rightarrow\begin{array}{cc}
\bar{m}=1 & \rightarrow\\
\bar{m}=0 & \rightarrow\\
\bar{m}=-1 & \rightarrow
\end{array}\mbox{\ensuremath{\frac{-1}{\sqrt{10}}}}\begin{array}{ccccc}
( & 0 & 0 & 0 & )\\
( & 0 & 0 & \sqrt{3} & )\\
( & 0 & \sqrt{3} & 0 & )
\end{array},
\end{equation}

\begin{equation}
m=-2\rightarrow\int d\widehat{\mathbf{p}}\mathbf{Z}_{2-2}^{\dagger}\left(\widehat{\mathbf{p}}\right)Y_{1\bar{m}}\left(\widehat{\mathbf{p}}\right)\Rightarrow\begin{array}{cc}
\bar{m}=1 & \rightarrow\\
\bar{m}=0 & \rightarrow\\
\bar{m}=-1 & \rightarrow
\end{array}\mbox{\ensuremath{\frac{-1}{\sqrt{5}}}}\begin{array}{ccccc}
( & 0 & 0 & 0 & )\\
( & 0 & 0 & 0 & )\\
( & 0 & 0 & \sqrt{3} & )
\end{array}.\label{eq:E_m2}
\end{equation}

Note that we list row vectors for each $\bar{m}=-1,0,1$. The summation
in $\bar{m}$ in Eq.~\ref{eq:E_m_2}-\ref{eq:E_m2} can be done and
by using Eq.~\ref{eq:J_and_r_def} and Eq.~\ref{eq:Y1m}, after
some simple algebra, we get:

\begin{eqnarray}
a_{22}^{\omega\,\,\bar{l}=1} & = & \frac{3ki}{\pi\sqrt{10}}\int d^{3}\mathbf{r}\left(J_{1}^{\omega}r_{1}\right)\frac{j_{1}\left(kr\right)}{kr},\nonumber \\
a_{21}^{\omega\,\,\bar{l}=1} & = & \frac{3ki}{\pi\sqrt{10}}\frac{1}{\sqrt{2}}\int d^{3}\mathbf{r}\left(J_{1}^{\omega}r_{0}+J_{0}^{\omega}r_{1}\right)\frac{j_{1}\left(kr\right)}{kr},\nonumber \\
a_{20}^{\omega\,\,\bar{l}=1} & = & \frac{3ki}{\pi\sqrt{10}}\frac{1}{\sqrt{6}}\int d^{3}\mathbf{r}\left(J_{1}^{\omega}r_{-1}+J_{-1}^{\omega}r_{1}+2J_{0}^{\omega}r_{0}\right)\frac{j_{1}\left(kr\right)}{kr},\nonumber \\
a_{2-1}^{\omega\,\,\bar{l}=1} & = & \frac{3ki}{\pi\sqrt{10}}\frac{1}{\sqrt{2}}\int d^{3}\mathbf{r}\left(J_{-1}^{\omega}r_{0}+J_{0}^{\omega}r_{-1}\right)\frac{j_{1}\left(kr\right)}{kr},\nonumber \\
a_{2-2}^{\omega\,\,\bar{l}=1} & = & \frac{3ki}{\pi\sqrt{10}}\int d^{3}\mathbf{r}\left(J_{-1}^{\omega}r_{-1}\right)\frac{j_{1}\left(kr\right)}{kr},\label{eq:a2m_l_1}
\end{eqnarray}

as the expression for the $\,a_{2m}^{\omega\,\,\bar{l}=1}$ contribution
of the electric quadrupole moment in spherical coordinate.

\subsubsection{$a_{2m}^{\omega\,\,\bar{l}=3}$ contribution}

In order to derive the $\,a_{2m}^{\omega\,\,\bar{l}=3}$ contribution
of the electric quadrupole moment in spherical coordinate, we use
Eq.~\ref{eq:EM}, i.e.

\begin{eqnarray}
a_{2m}^{\omega} & = & -\frac{4\pi i}{\sqrt{\left(2\pi\right)^{2}}}\stackrel[\overline{m}=-3]{3}{\sum}\int d\widehat{\mathbf{p}}\mathbf{Z}_{2m}^{\dagger}\left(\widehat{\mathbf{p}}\right)Y_{3\overline{m}}\left(\widehat{\mathbf{p}}\right)\int d^{3}\mathbf{r}\mathbf{J\left(r\right)}Y_{3\overline{m}}^{*}\left(\widehat{\mathbf{r}}\right)j_{3}\left(kr\right),
\end{eqnarray}

using Eq.~\ref{eq:X_Def} and Eq.~\ref{eq:Zjm}, we compute the
momentum integrals ($\int d\widehat{\mathbf{p}}\mathbf{Z}_{2m}^{\dagger}\left(\widehat{\mathbf{p}}\right)Y_{3\bar{m}}\left(\widehat{\mathbf{p}}\right)$)
for each $m$ case

\begin{equation}
m=2\rightarrow\int d\widehat{\mathbf{p}}\mathbf{Z}_{22}^{\dagger}\left(\widehat{\mathbf{p}}\right)Y_{3\bar{m}}\left(\widehat{\mathbf{p}}\right)\Rightarrow\begin{array}{cc}
\bar{m}=3 & \rightarrow\\
\bar{m}=2 & \rightarrow\\
\bar{m}=1 & \rightarrow\\
\bar{m}=0 & \rightarrow\\
\bar{m}=-1 & \rightarrow\\
\bar{m}=-2 & \rightarrow\\
\bar{m}=-3 & \rightarrow
\end{array}\mbox{\ensuremath{\frac{-1}{\sqrt{105}}}}\begin{array}{ccccc}
( & 0 & 0 & \sqrt{30} & )\\
( & 0 & -\sqrt{10} & 0 & )\\
( & \sqrt{2} & 0 & 0 & )\\
( & 0 & 0 & 0 & ),\\
( & 0 & 0 & 0 & )\\
( & 0 & 0 & 0 & )\\
( & 0 & 0 & 0 & )
\end{array}\label{eq:ZY_int_m2}
\end{equation}

\begin{equation}
m=1\rightarrow\int d\widehat{\mathbf{p}}\mathbf{Z}_{21}^{\dagger}\left(\widehat{\mathbf{p}}\right)Y_{3\bar{m}}\left(\widehat{\mathbf{p}}\right)\Rightarrow\begin{array}{cc}
\bar{m}=3 & \rightarrow\\
\bar{m}=2 & \rightarrow\\
\bar{m}=1 & \rightarrow\\
\bar{m}=0 & \rightarrow\\
\bar{m}=-1 & \rightarrow\\
\bar{m}=-2 & \rightarrow\\
\bar{m}=-3 & \rightarrow
\end{array}\mbox{\ensuremath{\frac{-1}{\sqrt{105}}}}\begin{array}{ccccc}
( & 0 & 0 & 0 & )\\
( & 0 & 0 & 2\sqrt{5} & )\\
( & 0 & -4 & 0 & )\\
( & \sqrt{6} & 0 & 0 & )\\
( & 0 & 0 & 0 & )\\
( & 0 & 0 & 0 & )\\
( & 0 & 0 & 0 & )
\end{array},
\end{equation}

\begin{equation}
m=0\rightarrow\int d\widehat{\mathbf{p}}\mathbf{Z}_{20}^{\dagger}\left(\widehat{\mathbf{p}}\right)Y_{3\bar{m}}\left(\widehat{\mathbf{p}}\right)\Rightarrow\begin{array}{cc}
\bar{m}=3 & \rightarrow\\
\bar{m}=2 & \rightarrow\\
\bar{m}=1 & \rightarrow\\
\bar{m}=0 & \rightarrow\\
\bar{m}=-1 & \rightarrow\\
\bar{m}=-2 & \rightarrow\\
\bar{m}=-3 & \rightarrow
\end{array}\mbox{\ensuremath{\frac{-1}{\sqrt{35}}}}\begin{array}{ccccc}
( & 0 & 0 & 0 & )\\
( & 0 & 0 & 0 & )\\
( & 0 & 0 & 2 & )\\
( & 0 & -\sqrt{6} & 0 & ),\\
( & 2 & 0 & 0 & )\\
( & 0 & 0 & 0 & )\\
( & 0 & 0 & 0 & )
\end{array}
\end{equation}

\begin{equation}
m=-1\rightarrow\int d\widehat{\mathbf{p}}\mathbf{Z}_{2-1}^{\dagger}\left(\widehat{\mathbf{p}}\right)Y_{3\bar{m}}\left(\widehat{\mathbf{p}}\right)\Rightarrow\begin{array}{cc}
\bar{m}=3 & \rightarrow\\
\bar{m}=2 & \rightarrow\\
\bar{m}=1 & \rightarrow\\
\bar{m}=0 & \rightarrow\\
\bar{m}=-1 & \rightarrow\\
\bar{m}=-2 & \rightarrow\\
\bar{m}=-3 & \rightarrow
\end{array}\mbox{\ensuremath{\frac{-1}{\sqrt{105}}}}\begin{array}{ccccc}
( & 0 & 0 & 0 & )\\
( & 0 & 0 & 0 & )\\
( & 0 & 0 & 0 & )\\
( & 0 & 0 & \sqrt{6} & ),\\
( & 0 & -4 & 0 & )\\
( & 2\sqrt{5} & 0 & 0 & )\\
( & 0 & 0 & 0 & )
\end{array}
\end{equation}

\begin{equation}
m=-2\rightarrow\int d\widehat{\mathbf{p}}\mathbf{Z}_{2-2}^{\dagger}\left(\widehat{\mathbf{p}}\right)Y_{3\bar{m}}\left(\widehat{\mathbf{p}}\right)\Rightarrow\begin{array}{cc}
\bar{m}=3 & \rightarrow\\
\bar{m}=2 & \rightarrow\\
\bar{m}=1 & \rightarrow\\
\bar{m}=0 & \rightarrow\\
\bar{m}=-1 & \rightarrow\\
\bar{m}=-2 & \rightarrow\\
\bar{m}=-3 & \rightarrow
\end{array}\mbox{\ensuremath{\frac{-1}{\sqrt{105}}}}\begin{array}{ccccc}
( & 0 & 0 & 0 & )\\
( & 0 & 0 & 0 & )\\
( & 0 & 0 & 0 & )\\
( & 0 & 0 & 0 & )\\
( & 0 & 0 & \sqrt{2} & )\\
( & 0 & -\sqrt{10} & 0 & )\\
( & \sqrt{30} & 0 & 0 & )
\end{array},\label{eq:ZY_int_m_2}
\end{equation}

Note that we list row vectors for each $\bar{m}=-3,-2,-1,0,1,2,3$.
The summation in $\bar{m}$ in Eq.~\ref{eq:ZY_int_m2}-\ref{eq:ZY_int_m_2}
can be done and by using Eq.~\ref{eq:J_and_r_def} and Eq.~\ref{eq:Y3m},
after some simple algebra, we get:

\begin{eqnarray}
a_{22}^{\omega\,\,\bar{l}=3} & = & \frac{-ik^{3}}{\pi\sqrt{10}}\int d^{3}\mathbf{r}\left[J_{1}^{\omega}\left(2r_{0}^{2}+r_{-1}r_{1}\right)r_{1}-5J_{0}^{\omega}r_{1}^{2}r_{0}+5J_{-1}^{\omega}r_{1}^{3}\right]\frac{j_{3}\left(kr\right)}{\left(kr\right)^{3}},\nonumber \\
a_{21}^{\omega\,\bar{\,l}=3} & = & \frac{-\sqrt{2}ik^{3}}{\pi\sqrt{10}}\int d^{3}\mathbf{r}\left[J_{1}^{\omega}\left(r_{0}^{3}+3r_{-1}r_{1}r_{0}\right)-2J_{0}^{\omega}\left(2r_{0}^{2}+r_{-1}r_{1}\right)r_{1}+5J_{-1}^{\omega}\left(r_{1}^{2}r_{0}\right)\right]\frac{j_{3}\left(kr\right)}{\left(kr\right)^{3}},\nonumber \\
a_{20}^{\omega\,\bar{\,l}=3} & = & \frac{-\sqrt{6}ik^{3}}{\pi\sqrt{10}}\int d^{3}\mathbf{r}\left[J_{1}^{\omega}\left(2r_{0}^{2}+r_{-1}r_{1}\right)r_{-1}-J_{0}^{\omega}\left(r_{0}^{3}+3r_{-1}r_{1}r_{0}\right)+J_{-1}^{\omega}\left(2r_{0}^{2}+r_{-1}r_{1}\right)r_{1}\right]\frac{j_{3}\left(kr\right)}{\left(kr\right)^{3}},\nonumber \\
a_{2-1}^{\omega\,\bar{\,l}=3} & = & \frac{-\sqrt{2}ik^{3}}{\pi\sqrt{10}}\int d^{3}\mathbf{r}\left[5J_{1}^{\omega}\left(r_{-1}^{2}r_{0}\right)-2J_{0}^{\omega}\left(2r_{0}^{2}+r_{-1}r_{1}\right)r_{-1}+J_{-1}^{\omega}\left(r_{0}^{3}+3r_{-1}r_{1}r_{0}\right)\right]\frac{j_{3}\left(kr\right)}{\left(kr\right)^{3}},\nonumber \\
a_{2-2}^{\omega\,\bar{\,l}=3} & = & \frac{-ik^{3}}{\pi\sqrt{10}}\int d^{3}\mathbf{r}\left[5J_{1}^{\omega}r_{-1}^{2}-5J_{0}^{\omega}r_{-1}r_{0}+J_{-1}^{\omega}\left(2r_{0}^{2}+r_{-1}r_{1}\right)\right]\hat{r}_{-1}\frac{j_{3}\left(kr\right)}{\left(kr\right)^{3}}.\label{eq:a2m_l3}
\end{eqnarray}

\subsection{Cartesian coordinates}

In this subsection, we introduce the exact electric quadrupole moment
in the Cartesian coordinates. As already highlighted the electric
quadrupole moment has two terms:

\begin{eqnarray*}
\mathbf{Q}^{e} & = & \mathbf{Q}^{e\,\,\bar{l}=1}+\mathbf{Q}^{e\,\,\bar{l}=3}.
\end{eqnarray*}

\subsubsection{$\mathbf{Q}^{e\,\,\bar{l}=1}$ contribution}

In order to obtain the Cartesian coordinates, we use the following
transformation between the spherical and Cartesian coordinates:

\begin{eqnarray}
Q_{xx}^{e\,\,\bar{l}=1} & = & C_{2}^{\mathrm{e}}\left[\frac{a_{22}^{\omega\,\,\bar{l}=1}+a_{2-2}^{\omega\,\,\bar{l}=1}}{2}-\frac{1}{\sqrt{6}}a_{20}^{\omega\,\,\bar{l}=1}\right],\nonumber \\
Q_{xy}^{e\,\,\bar{l}=1}=Q_{yx}^{e\,\,\bar{l}=1} & = & C_{2}^{\mathrm{e}}\left[\frac{a_{2-2}^{\omega\,\,\bar{l}=1}-a_{22}^{\omega\,\,\bar{l}=1}}{2i}\right],\nonumber \\
Q_{xz}^{e\,\,\bar{l}=1}=Q_{zx}^{e\,\,\bar{l}=1} & = & C_{2}^{\mathrm{e}}\left[\frac{a_{2-1}^{\omega\,\,\bar{l}=1}-a_{21}^{\omega\,\,\bar{l}=1}}{2}\right],\nonumber \\
Q_{yz}^{e\,\,\bar{l}=1}=Q_{zy}^{e\,\,\bar{l}=1} & = & C_{2}^{\mathrm{e}}\left[\frac{a_{2-1}^{\omega\,\,\bar{l}=1}+a_{21}^{\omega\,\,\bar{l}=1}}{2i}\right],\nonumber \\
Q_{yy}^{e\,\,\bar{l}=1} & = & C_{2}^{\mathrm{e}}\left[-\left(\frac{a_{22}^{\omega\,\,\bar{l}=1}+a_{2-2}^{\omega\,\,\bar{l}=1}}{2}\right)-\frac{1}{\sqrt{6}}a_{20}^{\omega\,\,\bar{l}=1}\right],\nonumber \\
Q_{zz}^{e\,\,\bar{l}=1} & = & C_{2}^{\mathrm{e}}\left[\frac{2}{\sqrt{6}}a_{20}^{\omega\,\,\bar{l}=1}\right],\label{eq:Qe_l_1}
\end{eqnarray}

where $C_{2}^{\mathrm{e}}=\frac{6\pi\sqrt{10}}{ck^{2}}$, which is
obtained by comparing the electric field expressions in spherical
\cite{Jackson:98} and Cartesian coordinates, i.e. Eq.~\ref{eq:E_Field}.
It is important to note that $\mathbf{Q}^{e}$ is a symmetric tensor
and traceless, i.e. $Q_{xx}^{e}+Q_{yy}^{e}+Q_{zz}^{e}=0$, which reduces
the Cartesian quadrupole electric moment to five independent components.

Now, by substitute Eq.~\ref{eq:a2m_l_1} in Eq.~\ref{eq:Qe_l_1},
and by using Eq.~\ref{eq:CP_S} and Eq.~\ref{eq:DP_C}. The final
results read as

\begin{eqnarray}
Q_{xx}^{e\,\,\bar{l}=1} & = & C_{2}^{\mathrm{e}}\left[\frac{a_{22}^{\omega\,\,\bar{l}=1}+a_{2-2}^{\omega\,\,\bar{l}=1}}{2}-\frac{1}{\sqrt{6}}a_{20}^{\omega\,\,\bar{l}=1}\right],\nonumber \\
 & = & C_{2}^{\mathrm{e}}\frac{3ki}{2\pi\sqrt{10}}\int d^{3}\mathbf{r}\left[\left(J_{-1}^{\omega}r_{-1}+J_{1}^{\omega}r_{1}\right)-\frac{1}{3}\left(J_{1}^{\omega}r_{-1}+J_{-1}^{\omega}r_{1}+2J_{0}^{\omega}r_{0}\right)\right]\frac{j_{1}\left(kr\right)}{kr},\nonumber \\
 & = & C_{2}^{\mathrm{e}}\frac{ki}{2\pi\sqrt{10}}\int d^{3}\mathbf{r}\left[3\left(J_{-1}^{\omega}r_{-1}+J_{1}^{\omega}r_{1}\right)-\left(J_{1}^{\omega}r_{-1}+J_{-1}^{\omega}r_{1}+2J_{0}^{\omega}r_{0}\right)\right]\frac{j_{1}\left(kr\right)}{kr},\nonumber \\
 & = & -\frac{3}{i\omega}\int d^{3}\mathbf{r}\left[3\left(2xJ_{x}^{\omega}-xJ_{x}^{\omega}-yJ_{y}^{\omega}\right)-\left(2zJ_{z}^{\omega}-xJ_{x}^{\omega}-yJ_{y}^{\omega}\right)\right]\frac{j_{1}\left(kr\right)}{kr},\nonumber \\
 & = & -\frac{3}{i\omega}\int d^{3}\mathbf{r}\left[3\left(xJ_{x}^{\omega}+xJ_{x}^{\omega}\right)-2\left(xJ_{x}^{\omega}+yJ_{y}^{\omega}+zJ_{z}^{\omega}\right)\right]\frac{j_{1}\left(kr\right)}{kr},\nonumber \\
 & = & -\frac{3}{i\omega}\int d^{3}\mathbf{r}\left[3\left(xJ_{x}^{\omega}+xJ_{x}^{\omega}\right)-2\left(\mathbf{r}\cdot\mathbf{J}_{\omega}\right)\right]\frac{j_{1}\left(kr\right)}{kr},
\end{eqnarray}

\begin{eqnarray}
Q_{xy}^{e\,\,\bar{l}=1} & = & C_{2}^{\mathrm{e}}\left[\frac{a_{2-2}^{\omega\,\,\bar{l}=1}-a_{22}^{\omega\,\,\bar{l}=1}}{2i}\right],\nonumber \\
 & = & C_{2}^{\mathrm{e}}\frac{ki}{2\pi\sqrt{10}}\int d^{3}\mathbf{r}\left(\frac{J_{-1}^{\omega}r_{-1}-J_{1}^{\omega}r_{1}}{i}\right)\frac{j_{1}\left(kr\right)}{kr},\nonumber \\
 & = & -\frac{3}{i\omega}\int d^{3}\mathbf{r}3\left(\frac{J_{-1}^{\omega}r_{-1}-J_{1}^{\omega}r_{1}}{i}\right)\frac{j_{1}\left(kr\right)}{kr},\nonumber \\
 & = & -\frac{3}{i\omega}\int d^{3}\mathbf{r}3\left(xJ_{y}^{\omega}+yJ_{x}^{\omega}\right)\frac{j_{1}\left(kr\right)}{kr},
\end{eqnarray}

\begin{eqnarray}
Q_{xz}^{e\,\,\bar{l}=1} & = & C_{2}^{\mathrm{e}}\left[\frac{a_{2-1}^{\omega\,\,\bar{l}=1}-a_{21}^{\omega\,\,\bar{l}=1}}{2}\right],\nonumber \\
 & = & C_{2}^{\mathrm{e}}\frac{ik}{2\pi\sqrt{10}}\int d^{3}\mathbf{r}3\sqrt{2}\left[J_{-1}^{\omega}r_{0}+J_{0}^{\omega}r_{-1}-J_{1}^{\omega}r_{0}-J_{0}^{\omega}r_{1}\right]\frac{j_{1}\left(kr\right)}{kr},\nonumber \\
 & = & -\frac{3}{i\omega}\int d^{3}\mathbf{r}3\sqrt{2}\left[\left(J_{-1}^{\omega}-J_{1}^{\omega}\right)r_{0}+\left(r_{-1}-r_{1}\right)J_{0}^{\omega}\right]\frac{j_{1}\left(kr\right)}{kr},\nonumber \\
 & = & -\frac{3}{i\omega}\int d^{3}\mathbf{r}3\left(xJ_{z}^{\omega}+zJ_{x}^{\omega}\right)\frac{j_{1}\left(kr\right)}{kr},
\end{eqnarray}

\begin{eqnarray}
Q_{yz}^{e\,\,\bar{l}=1} & = & C_{2}^{\mathrm{e}}\left[\frac{a_{2-1}^{\omega\,\,\bar{l}=1}+a_{21}^{\omega\,\,\bar{l}=1}}{2i}\right],\nonumber \\
 & = & C_{2}^{\mathrm{e}}\frac{ki}{2\pi\sqrt{10}}\int d^{3}\mathbf{r}\frac{3}{\sqrt{2}i}\left(J_{-1}^{\omega}r_{0}+J_{0}^{\omega}r_{-1}+J_{1}^{\omega}r_{0}+J_{0}^{\omega}r_{1}\right)\frac{j_{1}\left(kr\right)}{kr},\nonumber \\
 & = & -\frac{3}{i\omega}\int d^{3}\mathbf{r}\frac{3}{\sqrt{2}i}\left(\left[J_{-1}^{\omega}+J_{1}^{\omega}\right]r_{0}+\left[r_{-1}+r_{1}\right]J_{0}^{\omega}\right)\frac{j_{1}\left(kr\right)}{kr},\nonumber \\
 & = & -\frac{3}{i\omega}\int d^{3}\mathbf{r}3\left(yJ_{z}^{\omega}+zJ_{y}^{\omega}\right)\frac{j_{1}\left(kr\right)}{kr},
\end{eqnarray}

\begin{eqnarray}
Q_{yy}^{e\,\,\bar{l}=1} & = & C_{2}^{\mathrm{e}}\left[-\left(\frac{a_{22}^{\omega\,\,\bar{l}=1}+a_{2-2}^{\omega\,\,\bar{l}=1}}{2}\right)-\frac{1}{\sqrt{6}}a_{20}^{\omega\,\,\bar{l}=1}\right],\nonumber \\
 & = & C_{2}^{\mathrm{e}}\frac{ki}{2\pi\sqrt{10}}\int d^{3}\mathbf{r}\left[-3\left(J_{-1}^{\omega}r_{-1}+J_{1}^{\omega}r_{1}\right)-\left(J_{1}^{\omega}r_{-1}+J_{-1}^{\omega}r_{1}+2J_{0}^{\omega}r_{0}\right)\right]\frac{j_{1}\left(kr\right)}{kr},\nonumber \\
 & = & -\frac{3}{i\omega}\int d^{3}\mathbf{r}\left[-3\left(J_{-1}^{\omega}r_{-1}+J_{1}^{\omega}r_{1}\right)-\left(J_{1}^{\omega}r_{-1}+J_{-1}^{\omega}r_{1}+2J_{0}^{\omega}r_{0}\right)\right]\frac{j_{1}\left(kr\right)}{kr},\nonumber \\
 & = & -\frac{3}{i\omega}\int d^{3}\mathbf{r}\left[3\left(2yJ_{y}^{\omega}-xJ_{x}^{\omega}-yJ_{y}^{\omega}\right)-\left(2zJ_{z}^{\omega}-xJ_{x}^{\omega}-yJ_{y}^{\omega}\right)\right]\frac{j_{1}\left(kr\right)}{kr},\nonumber \\
 & = & -\frac{3}{i\omega}\int d^{3}\mathbf{r}\left[3\left(yJ_{y}^{\omega}+yJ_{y}^{\omega}\right)-2\left(xJ_{x}^{\omega}+yJ_{y}^{\omega}+zJ_{z}^{\omega}\right)\right]\frac{j_{1}\left(kr\right)}{kr},\nonumber \\
 & = & -\frac{3}{i\omega}\int d^{3}\mathbf{r}\left[3\left(yJ_{y}^{\omega}+yJ_{y}^{\omega}\right)-2\left(\mathbf{r}\cdot\mathbf{J}_{\omega}\right)\right]\frac{j_{1}\left(kr\right)}{kr},
\end{eqnarray}

\begin{eqnarray}
Q_{zz}^{e\,\,\bar{l}=1} & = & C_{2}^{\mathrm{e}}\left[\frac{2}{\sqrt{6}}a_{20}^{\omega\,\,\bar{l}=1}\right]\nonumber \\
 & = & C_{2}^{\mathrm{e}}\frac{ki}{2\pi\sqrt{10}}\int d^{3}\mathbf{r}2\left[J_{1}^{\omega}r_{-1}+J_{-1}^{\omega}r_{1}+2J_{0}^{\omega}r_{0}\right]\frac{j_{1}\left(kr\right)}{kr},\nonumber \\
 & = & -\frac{3}{i\omega}\int d^{3}\mathbf{r}2\left[J_{1}^{\omega}r_{-1}+J_{-1}^{\omega}r_{1}+2J_{0}^{\omega}r_{0}\right]\frac{j_{1}\left(kr\right)}{kr},\nonumber \\
 & = & -\frac{3}{i\omega}\int d^{3}\mathbf{r}2\left[zJ_{z}^{\omega}-xJ_{x}^{\omega}-yJ_{y}^{\omega}\right]\frac{j_{1}\left(kr\right)}{kr},\nonumber \\
 & = & -\frac{3}{i\omega}\int d^{3}\mathbf{r}2\left[3zJ_{z}^{\omega}-\left(xJ_{x}^{\omega}+yJ_{y}^{\omega}+zJ_{z}^{\omega}\right)\right]\frac{j_{1}\left(kr\right)}{kr},\nonumber \\
 & = & -\frac{3}{i\omega}\int d^{3}\mathbf{r}\left[3\left(zJ_{z}^{\omega}+zJ_{z}^{\omega}\right)-2\left(xJ_{x}^{\omega}+yJ_{y}^{\omega}+zJ_{z}^{\omega}\right)\right]\frac{j_{1}\left(kr\right)}{kr},\nonumber \\
 & = & -\frac{3}{i\omega}\int d^{3}\mathbf{r}\left[3\left(zJ_{z}^{\omega}+zJ_{z}^{\omega}\right)-2\left(\mathbf{r}\cdot\mathbf{J}_{\omega}\right)\right]\frac{j_{1}\left(kr\right)}{kr},
\end{eqnarray}

in the above derivations, we used Eq.~\ref{eq:CP_S} , Eq.\ref{eq:DP_C}
and Eq.~\ref{eq:r_C2S}. Finally, we can write these above expressions
for magnetic quadrupole in a short form:

\[
\boxed{Q_{\alpha\beta}^{e\,\,\bar{l}=1}=-\frac{3}{i\omega}\int d^{3}\mathbf{r}\left[3\left(r_{\beta}J_{\alpha}^{\omega}+r_{\alpha}J_{\beta}^{\omega}\right)-2\left(\mathbf{r}\cdot\mathbf{J}_{\omega}\right)\delta_{\alpha\beta}\right]\frac{j_{1}\left(kr\right)}{kr}}
\]

\subsubsection{$\mathbf{Q}^{e\,\,\bar{l}=3}$ contribution}

In this subsection, we obtain the electric quadrupole moment in Cartesian
coordinate only for one of the components $\mathbf{Q}^{e\,\,\bar{l}=3}$,
i.e.

\begin{eqnarray}
Q_{zz}^{e\,\bar{\,l}=3} & = & C_{2}^{\mathrm{e}}\left[\frac{2}{\sqrt{6}}a_{20}^{\omega\,\bar{\,l}=3}\right],\nonumber \\
 & = & C_{2}^{\mathrm{e}}\frac{ik^{3}}{\pi\sqrt{10}}\int d^{3}\mathbf{r}2\left[-J_{1}^{\omega}\left(2r_{0}^{2}+r_{-1}r_{1}\right)r_{-1}+J_{0}^{\omega}\left(r_{0}^{3}+3r_{-1}r_{1}r_{0}\right)-J_{-1}^{\omega}\left(2r_{0}^{2}+r_{-1}r_{1}\right)r_{1}\right]\frac{j_{3}\left(kr\right)}{\left(kr\right)^{3}},\nonumber \\
 & = & -\frac{6k^{2}}{i\omega}\int d^{3}\mathbf{r}\left[5zz\left(xJ_{x}^{\omega}+yJ_{y}^{\omega}+zJ_{z}^{\omega}\right)-\left(zJ_{z}^{\omega}+zJ_{z}^{\omega}\right)r^{2}-r^{2}\left(xJ_{x}^{\omega}+yJ_{y}^{\omega}+zJ_{z}^{\omega}\right)\right]\frac{j_{3}\left(kr\right)}{\left(kr\right)^{3}},
\end{eqnarray}

similar expressions can be obtained for the other components. The
short form of the above expression read as

\[
\boxed{Q_{\alpha\beta}^{e\,\,\bar{l}=3}=-\frac{6k^{2}}{i\omega}\int d^{3}\mathbf{r}\left[5r_{\alpha}r_{\beta}\left(\mathbf{r}\cdot\mathbf{J}_{\omega}\right)-\left(r_{\alpha}J_{\beta}+r_{\beta}J_{\alpha}\right)r^{2}-r^{2}\left(\mathbf{r}\cdot\mathbf{J}_{\omega}\right)\delta_{\alpha\beta}\right]\frac{j_{3}\left(kr\right)}{\left(kr\right)^{3}}}
\]

where $\alpha,\beta=x,y,z$.

\subsubsection{$\mathbf{Q}^{e\,\,\bar{l}=1}$ and $\mathbf{Q}^{e\,\,\bar{l}=3}$
contribution}

Finally, the total electric quadrupole moment in the Cartesian coordinates
can be obtained:

\begin{eqnarray}
Q_{\alpha\beta}^{\mathrm{e}} & = & Q_{\alpha\beta}^{e\,\,\bar{l}=1}+Q_{\alpha\beta}^{e\,\,\bar{l}=3},\\
 &  & -\frac{3}{i\omega}\left\{ \int d^{3}\mathbf{r}\left[3\left(r_{\beta}J_{\alpha}^{\omega}+r_{\alpha}J_{\beta}^{\omega}\right)-2\left(\mathbf{r}\cdot\mathbf{J}_{\omega}\right)\delta_{\alpha\beta}\right]\frac{j_{1}\left(kr\right)}{kr}\right.\nonumber \\
 &  & \left.+2k^{2}\int d^{3}\mathbf{r}\left[5r_{\alpha}r_{\beta}\left(\mathbf{r}\cdot\mathbf{J}_{\omega}\right)-\left(r_{\alpha}J_{\beta}+r_{\beta}J_{\alpha}\right)r^{2}-r^{2}\left(\mathbf{r}\cdot\mathbf{J}_{\omega}\right)\delta_{\alpha\beta}\right]\frac{j_{3}\left(kr\right)}{\left(kr\right)^{3}}\right\} .\nonumber
\end{eqnarray}

This expression is documented in Tab. 2 of the main manuscript.

\subsection{Long-wavelength approximation }

We now can make the \textit{long-wavelength approximation} by using
the small argument approximation to the spherical Bessel function,
i.e.

\begin{eqnarray}
j_{1}\left(kr\right) & \approx & \frac{kr}{3}\left(1-\frac{\left(kr\right)^{2}}{10}\right),\nonumber \\
j_{3}\left(kr\right) & \approx & \frac{\left(kr\right)^{3}}{105},
\end{eqnarray}

and obtain the well-known long-wavelength expression for the electric
dipole moments:

\begin{eqnarray}
Q_{\alpha\beta}^{\mathrm{e}} & = & Q_{\alpha\beta}^{e\,\,\bar{l}=1}+Q_{\alpha\beta}^{e\,\,\bar{l}=3},\\
 &  & -\frac{3}{i\omega}\left\{ \int d^{3}\mathbf{r}\left[3\left(r_{\beta}J_{\alpha}^{\omega}+r_{\alpha}J_{\beta}^{\omega}\right)-2\left(\mathbf{r}\cdot\mathbf{J}_{\omega}\right)\delta_{\alpha\beta}\right]\frac{j_{1}\left(kr\right)}{kr}\right.\nonumber \\
 &  & \left.+2k^{2}\int d^{3}\mathbf{r}\left[5r_{\alpha}r_{\beta}\left(\mathbf{r}\cdot\mathbf{J}_{\omega}\right)-\left(r_{\alpha}J_{\beta}+r_{\beta}J_{\alpha}\right)r^{2}-r^{2}\left(\mathbf{r}\cdot\mathbf{J}_{\omega}\right)\delta_{\alpha\beta}\right]\frac{j_{3}\left(kr\right)}{\left(kr\right)^{3}}\right\} ,\nonumber \\
 & \approx & \frac{1}{-i\omega}\left\{ \int d^{3}\mathbf{r}\left[3\left(r_{\beta}J_{\alpha}^{\omega}+r_{\alpha}J_{\beta}^{\omega}\right)-2\left(\mathbf{r}\cdot\mathbf{J}_{\omega}\right)\delta_{\alpha\beta}\right]\right.\nonumber \\
 &  & \left.+k^{2}\left(\frac{1}{14}\int d^{3}\mathbf{r}\left[4r_{\alpha}r_{\beta}\left(\mathbf{r}\cdot\mathbf{J}_{\omega}\right)-5r^{2}\left(r_{\alpha}J_{\beta}+r_{\beta}J_{\alpha}\right)+2r^{2}\left(\mathbf{r}\cdot\mathbf{J}_{\omega}\right)\delta_{\alpha\beta}\right]\right)\right\} .\nonumber
\end{eqnarray}

This expression is documented in Tab. 1 of the main manuscript. It
is important to note that the first term, i.e.

\begin{equation}
-\frac{1}{i\omega}\int d^{3}\mathbf{r}\left[3\left(r_{\beta}J_{\alpha}^{\omega}+r_{\alpha}J_{\beta}^{\omega}\right)-2\left(\mathbf{r}\cdot\mathbf{J}_{\omega}\right)\delta_{\alpha\beta}\right],
\end{equation}

can be found in \textit{any} electrodynamics textbook~\cite{Jackson:98}.
It is often considered as the \textit{electric quadrupole momen}t.
However, we shall always keep in mind that this expression is \textit{only}
valid for very small particles compared to wavelength. The second
term

\begin{equation}
\frac{1}{-i\omega}\left\{ k^{2}\left(\frac{1}{14}\int d^{3}\mathbf{r}\left[4r_{\alpha}r_{\beta}\left(\mathbf{r}\cdot\mathbf{J}_{\omega}\right)-5r^{2}\left(r_{\alpha}J_{\beta}+r_{\beta}J_{\alpha}\right)+2r^{2}\left(\mathbf{r}\cdot\mathbf{J}_{\omega}\right)\delta_{\alpha\beta}\right]\right)\right\} ,
\end{equation}

sometimes called the \textit{toroidal} quadrupole moment. They have
been \textit{incorrectly} called the \textit{third} family of multipole
moments. However, based on our derivation (see also Ref.~\cite{FerCor2015b}),
it is obvious that both terms are belongs to the electric quadrupole
moment.

\section{Radiation fields}

In this section, we wish to highlight the relation between the exact
induced multipole moments and radiated electric field~\cite{Jackson:98}:

\begin{eqnarray}
\mathbf{E} & = & \frac{k^{2}}{4\pi\epsilon_{0}}\frac{e^{ikr}}{r}\left\{ \mathbf{n}\times\left(\mathbf{p}\times\mathbf{n}\right)+\frac{1}{c}\left(\mathbf{m}\times\mathbf{n}\right)-\frac{ik}{6}\mathbf{n}\times\left(\mathbf{Q^{e}}\times\mathbf{n}\right)-\frac{ik}{6c}\left(\mathbf{Q}^{m}\times\mathbf{n}\right)\right\} .\label{eq:E_Field}
\end{eqnarray}

\end{widetext}

\end{document}